\documentclass{emulateapj}
\usepackage[english]{babel}     
\usepackage{enumitem}
\usepackage{amssymb,amsmath}    
\usepackage{graphicx}                   
\usepackage{braket}             
\usepackage{hyperref}
\usepackage{natbib}
\usepackage{bm}

\slugcomment{First submitted to ApJ: 14 October, 2015.}

\shorttitle{LEDA 087300}
\shortauthors{Graham, Ciambur \& Soria}

\begin{document}

\title{Does the intermediate mass black hole in LEDA~87300 (RGG~118) follow the
  near-quadratic $M_{\rm bh}$--$M_{\rm spheroid}$ relation?} 

\author{Alister W.\ Graham, Bogdan C.\ Ciambur} 
\affil{Centre for Astrophysics and Supercomputing, Swinburne University of
  Technology, Victoria 3122, Australia.}
\email{AGraham@astro.swin.edu.au}
\and
\author{Roberto Soria}
\affil{International Centre for Radio Astronomy Research, Curtin University,
  GPO Box U1987, Perth, WA 6845, Australia.}

\begin{abstract}

The mass scaling relation between supermassive black holes and their host
spheroids has previously been described by a quadratic or steeper relation at
low masses ($10^5 < M_{\rm bh}/M_{\odot} \lesssim 10^7$).  How this extends
into the realm of intermediate mass black holes ($10^2 < M_{\rm bh}/M_{\odot}
< 10^5$) is not yet clear, although for the barred Sm galaxy LEDA~87300,
Baldassare et al.\ have recently reported a nominal virial mass $M_{\rm
  bh}=5\times10^4~M_{\odot}$ residing in a `spheroid' of stellar mass equal to
$6.3\times10^8~M_{\odot}$.  We point out, for the first time, that LEDA~87300
therefore appears to reside on the near-quadratic $M_{\rm bh}$--$M_{\rm
  sph,*}$ relation.  However, Baldassare et al.\ modelled the bulge {\it and}
bar as the single spheroidal component of this galaxy.  Here we perform a
3-component bulge$+$bar$+$disk decomposition and find a bulge luminosity which
is 7.7 times fainter than the published `bulge' luminosity.  
After correcting for dust, we find that $M_{\rm
  bulge}=0.9\times10^8~M_{\odot}$ and $M_{\rm bulge}/M_{\rm disk}=0.04$ ---
which is now in accord with ratios typically found in Scd--Sm galaxies.  We go
on to discuss slight revisions to the stellar velocity dispersion ($40\pm11$
km s$^{-1}$) and black hole mass ($M_{\rm
  bh}=2.9^{+6.7}_{-2.3}\times10^4\,f_{2.3}~M_{\odot}$) and show that
LEDA~87300 remains consistent with the $M_{\rm bh}$--$\sigma$ relation, 
and also the near-quadratic $M_{\rm bh}$--$M_{\rm sph,*}$ relation when using the
reduced bulge mass.  LEDA~87300 therefore offers the first support for the rapid but
regulated (near-quadratic) growth of black holes, relative to their host
bulge/spheroid, extending into the domain of intermediate mass black holes.

\end{abstract}


\section{Introduction}

Although stellar mass black holes ($< 10^2 ~M_{\odot}$,
e.g., Belczynski et al.\ 2010) and supermassive black holes ($> 10^5
~M_{\odot}$) are now known entities (as reviewed in Kormendy \& Ho 2013 and 
Graham 2015a), 
there remains a paucity of intermediate mass black holes (IMBHs).  While there
are well over 100 galaxies believed to contain black holes with masses in the range
$10^5 < (M_{\rm bh}\, f_3)/M_{\odot} < 10^6$ (e.g., Greene \& Ho 2004; Jiang
et al.\ 2011; Xiao et al.\ 2011; Graham \& Scott 2015)\footnote{We use the
  notation $f_3 = f/3$ to denote that a virial factor $f=3$ has been
  adopted. If the actual virial factor $f$ is greater or smaller than 3, then
  the black hole masses need to be multiplied by $f/3$. The notation $f_4$
  indicates that a virial factor $f=4$ was used to derive the black hole
  mass.}, few IMBH candidates are currently known\footnote{They tend to be 
  referred to as `candidates' because our scientific instruments typically do not have
  the ability to spatially resolve the Keplerian orbits of the gas and stars 
  under the gravitational dominance of such black holes in other galaxies. 
  There also remains uncertainty as to the nature and mass of many of the 
  ultraluminous X-ray sources which might (not) be IMBHs 
  (e.g., Feng \& Soria 2011; Liu et al.\ 2013; Ek{\c s}i et al.\ 2015;
  Fabrika et al.\ 2015; Lasota et al.\ 2015; Zhou 2015).}.  
However, a battery of indirect methods have been 
used to imply their existence.  For example, perhaps the best such candidate
is the hyperluminous X-ray source HLX-1 ($M_{\rm bh} = 0.3$--$30\times 10^4
~M_{\odot}$), situated off-center in the lenticular galaxy ESO~243-49 (Farrell
et al.\ 2009, 2014; Soria et al.\ 2010; Webb et al.\ 2010, 2012, 2014; 
Davis et al.\ 2011; Cseh et al.\ 2015).

There are three groups of sources touted as IMBH candidates, namely: {\it (i)}
some highly-luminous, off-center, accreting sources in galaxies, such as HLX-1
in ESO~243-49 (these hyperluminous X-ray sources are special cases among the
so called ultraluminous X-ray sources)\footnote{Ultraluminous X-ray sources
  have $L_X > 3\times 10^{39}$ ergs, while hyperluminous X-ray sources have
  $L_X > 10^{41}$ ergs (Gao et al.\ 2003).}; {\it (ii)} centrally-located
black holes in dwarf and late-type spiral galaxies (such as LEDA~87300), and
{\it (iii)} (quiescent) black hole candidates in massive globular clusters
which have been proposed from dynamical modelling (Wyller 1970; Lutzgendorf et
al.\ 2013, and references therein) but not yet convincingly detected (e.g.,
Lanzoni 2015 and references therein).  The three groups may or may not have a
distinct physical origin. For example, a source that appears as an off-center
hyperluminous X-ray source today may have been the nucleus of an accreted and
stripped dwarf satellite galaxy (e.g., Drinkwater et al.\ 2003; Seth et
al.\ 2014), or perhaps an ejected nucleus (e.g., Gualandris \& Merritt 2008;
Komossa \& Merritt 2008; Merritt et al.\ 2009).  NGC~2276-3c is one such
off-center X-ray source and probable IMBH, with a reported mass of
$\sim$$5\times10^4 ~M_{\odot}$ (Mezcua et al.\ 2015a).  Another off-center
example can be found in the bulgeless disk galaxy NGC~4178, with a reported
black hole mass of $0.6\pm0.2 \times 10^4 ~M_{\odot}$; and the center of this
galaxy may harbour another black hole, in the mass range $10^4$--$10^5
~M_{\odot}$ (Secrest et al.\ 2012).  The non-central, ultraluminous X-ray
sources in NGC~1313 (X-2, Liu et al.\ 2012 and Pasham et al.\ 2014; and X-1,
Colbert \& Mushotzky 1999, Miller et al.\ 2013, and Pasham et al.\ 2015), and
in Zwicky~18 (Kaaret \& Feng 2013) may also signal IMBHs, wih further examples
given in Sutton et al.\ (2012).  If such IMBHs (see also Oka et al.\ 2015)
migrate to the centers of 
bulges through dynamical friction, it stands to reason that they may
contribute to the apparent bulge-(black hole) connection if the inward bound
IMBH is of significant mass compared to any black hole mass currently at the
center of the galaxy in question.  Thus, it is thought that the
centrally-located, intermediate and supermassive black holes may not
necessarily originate from a single seed, but rather many.

While a few IMBH candidates are known, as noted above, the galaxy LEDA~87300
(referred to as ``RGG~118'' by Baldassare et al.\ 2015; hereafter
BRGG15)\footnote{These authors renamed LEDA~87300 after the initials of their
  previous paper which had included this galaxy in their sample: Reines,
  Greene \& Geha (2013).}, hosts a particularly interesting IMBH candidate
because it is centrally located in this galaxy's bulge.  This galaxy can
therefore be used to probe the low-mass end of the (black hole mass)-(host
bulge) diagrams and scaling relations.  These include (i) the $M_{\rm
  bh}$--$\sigma$ relation (Ferrarese \& Merritt 2000; Gebhardt et al.\ 2000;
Park et al.\ 2012; Kormendy \& Ho 2013; McConnell \& Ma 2013; 
Sabra et al.\ 2015; Savorgnan \& Graham 2015c) 
and the offset nature of barred galaxies (Graham 2008; Hu 2008; Graham et
al.\ 2011), (ii) the 
near-linear $M_{\rm bh}$--$M_{\rm sph}$ relation\footnote{The host spheroid's 
  stellar mass, $M_{\rm sph,*}$, and dynamical mass, $M_{\rm sph,dyn}$, have
  been used in different studies.} (Magorrian et al.\ 1998;
McLure \& Dunlop 2002; Marconi \& Hunt 2003; H\"aring \& Rix 2004; 
Sani et al.\ 2011; Vika et al.\ 2012; Beifiori et al.\ 2012; L\"asker et al.\ 2014b) and
importantly the steeper $M_{\rm bh}$--$M_{\rm sph}$ relation at low masses 
(Graham \& Scott 2013, 2015; 
Savorgnan et al.\ 2015; see also Laor 1998, 2001), and (iii) the $M_{\rm bh}$--$n$
relation (Graham et al.\ 2001; Savorgnan et al.\ 2013)
involving the S\'ersic (1963) index of the host spheroid, i.e.\ the radial
distribution of stars.

The nature of these black hole scaling relations at low masses is not only of
interest for probing into the realm of the largely `missing' population of
IMBHs, but is expected to provide insight into the seed black holes that grew
into supermassive black holes (e.g., Madau \& Rees 2001; Miller \& Hamilton
2002; Portegies Zwart \& McMillan 2002; Johnson et al.\ 2013; Latif et
al.\ 2013).
For example, Alexander \& Natarajan (2014), Madau et al.\ (2014) and 
Lupi et al.\ (2015) illustrate how stellar mass black holes can rapidly grow
into $10^4 M_{\odot}$ black holes via radiatively inefficient super-critical
accretion, bypassing studies which {\it start} with initial seed black holes
with masses $\sim$$10^5 M_{\odot}$.
While Kormendy \& Ho (2013) advocate that black 
holes in low-mass (pseudo)bulges are randomly offset to the lower right of the
$M_{\rm bh}$--$M_{\rm sph}$ relations defined by high mass
``classical'' bulges --- an idea introduced by Graham (2008) and Hu (2008) --- 
Graham (2012, 2015a) has since shown that a steeper than linear relation 
exists for pseudobulges and low-mass classical bulges alike. 
In particular, a near-quadratic relation was found for the S\'ersic spheroids
($M_{\rm bh} \propto M_{\rm sph,*}^{2.22\pm0.58}$) while a near-linear
relation was found for the core-S\'ersic spheroids ($M_{\rm bh} \propto M_{\rm
  sph,*}^{0.97\pm0.14}$).  Savorgnan et al.\ (2015) have since revealed that
an alternative division may be between the early-type (red sequence) galaxies (with 
$M_{\rm bh} \propto M_{\rm sph,*}^{1.04\pm0.10}$) and the late-type
(blue sequence) galaxies which define a relation 
$M_{\rm bh} \propto M_{\rm sph,*}^{2}--M_{\rm sph,*}^{3}$. 
Using virial black hole masses, the bend in the $M_{\rm bh}$--$M_{\rm sph}$ relation
can now be seen all the way down to $M_{\rm bh} = 10^5 ~M_{\odot}$ (Graham \&
Scott 2015).  Spheroids with IMBH candidates are needed to test if this steep
relation continues to yet lower masses.  

\begin{figure}
\includegraphics[trim=3cm 1cm 8.5cm 1.35cm, width=\columnwidth]{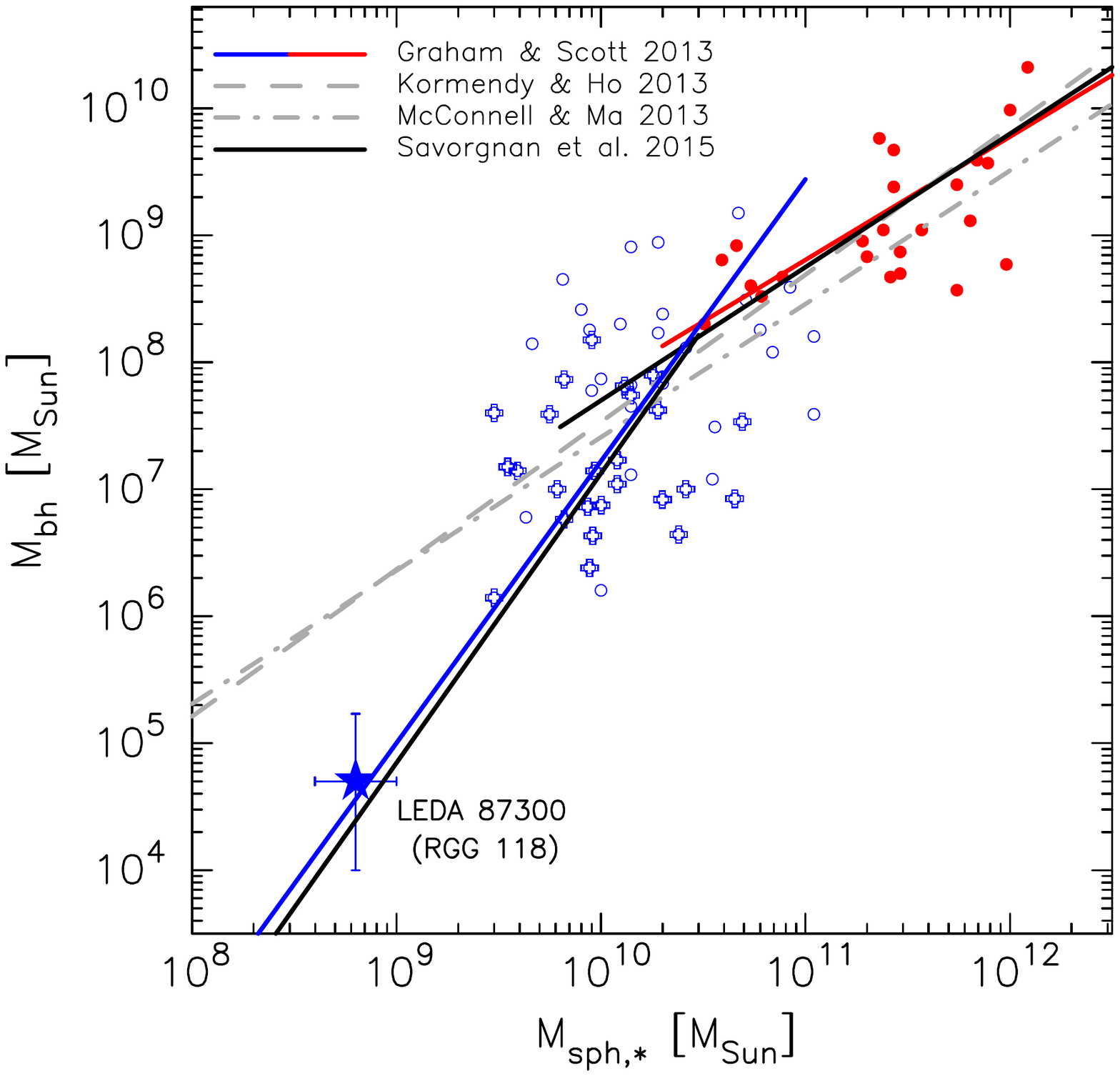}
\caption{Location of LEDA~87300 in the $M_{\rm bh}$--$M_{\rm sph,*}$ diagram
  according to Baldassare et al.\ (2015).  The near-quadratic relation from
  Scott et al.\ (2013) for S\'ersic galaxies (blue points) is shown in blue,
  and for core-S\'ersic galaxies (red points) in red. Barred galaxies are
  denoted by the crosses.  The near-quadratic relation from Savorgnan et al.\ 
   (2015) for spiral galaxies is shown in black (derived using the modified
  FITEXY routine from Tremaine et al.\ 2002 in a symmetrical manner) along
  with Savorgnan et al.'s near-linear relation for early-type galaxies. }
\label{Fig1}
\end{figure}

BRGG15 did not explore whether the $5\times10^4 f_4~M_{\odot}$ black hole in 
LEDA~87300 agrees (or not) with the steep $M_{\rm bh}$--$M_{\rm sph,*}$ relation.  
Figure~\ref{Fig1} shows the location of LEDA~87300 in the $M_{\rm
  bh}$--$M_{\rm sph,*}$ diagram according to the masses derived by
BRGG15 for LEDA~87300.  Importantly, it additionally shows, for the first time, where this IMBH
resides relative to the extrapolation to low masses of the near-quadratic
scaling relation constructed at higher masses.  The agreement is striking.  
This is important because, if confirmed, it suggests that we can 
predict the masses of (at least the more massive) IMBHs. 

It is obviously important to perform a correct decomposition of a galaxy's
light if we are to obtain the physical properties of the spheroidal component
and properly explore the (black hole)-(host spheroid) connection.  This is
true at both ends of the mass spectrum.  For example, isolating the spheroid
light from that of the galaxy has recently resulted in the discovery of the
missing population of compact, massive spheroids observed at high-redshifts
(referred to as red nuggets by Damjanov et al.\ 2009) 
but previously thought to be absent in the nearby Universe (Graham, Dullo \& Savorgnan 2015). 
At the same time, other studies which fit {\it large-scale} disks to some
of these nearby galaxies which actually contain {\it
  intermediate-scale} disks has resulted in the under-estimation of the spheroid
mass and thus the erroneous claims of overly high $M_{\rm bh}/M_{\rm sph,*}$ ratios, as
explained in Savorgnan \& Graham (2015b).

Here we provide an independent decomposition for the barred galaxy LEDA~87300,
to check if its low-mass spheroidal component follows the near-quadratic
$M_{\rm bh}$--$M_{\rm sph,*}$ relation as seen in Figure~\ref{Fig1}.  Although
BRGG15 performed a bulge/disk decomposition, they reported a bulge-to-disk
ratio of $\sim$0.3.  This is unusually high for an Sm galaxy (see Graham \&
Worley 2008), and appears to have arisen because they 
lumped the bar plus bulge light together, fitting for
what we call the ``barge''.  This also means that the S\'ersic index for the
bulge component of this galaxy is yet to be measured.

In Section~\ref{Sec_Data} we present the imaging data and our decomposition
analysis, refitting for both the ``barge''+disk and the bar+bulge+disk.  To be
thorough, in Section~3 we additionally revisit the derived black hole mass,
the spheroid velocity dispersion, and the X-ray data which suggested the
presence of an active galactic nucleus (AGN). Specifically, in
subsection~\ref{Sec_BHmass} we revise the nominal black hole mass using the
optimal virial $f$-factor for barred galaxies.  In subsection~\ref{Sec_sigma}
we provide an estimate of the stellar rather than the gas velocity dispersion,
and in subsection~\ref{Sec_X} we report on our confirmation of the X-ray
detection made by BRGG15.  In Section~\ref{Sec_DC} we present our revised
location of LEDA~87300 in the $M_{\rm bh}$--$M_{\rm sph,*}$, $M_{\rm
  bh}$--$\sigma$, and $M_{\rm bh}$--$n$ diagrams.  We provide a brief
discussion of its location in these diagrams and comment on potential future
targets.

BRGG15 did not identify ``RGG118'' with any pre-existing galaxy, and we
therefore provide some additional references and information here. 
LEDA~87300 is located at R.A.\ = 15 23 05.0, Dec = +11 45 53 (J2000:
Paturel et al.\ 2000) and is also known as PGC~87300; Paturel et
al.\ 1989)\footnote{The Catalogue of Principal Galaxies.}.  
BRGG15 have reported a virial mass of $5\times10^4 f_4 ~M_{\odot}$, with a
(1-sigma) range from (1--17)$\times 10^4 \, f_4 ~M_{\odot}$, for the black
hole located at the center of this catalogued low surface
brightness\footnote{``Low surface brightness galaxies'' have a central disk
  surface brightness more than 1 mag arcsec$^{-2}$ fainter than the canonical
  Freeman (1970) value of 21.65 B mag arcsec$^{-2}$.}, Sm
galaxy (LSBC F725-V01; Schombert et al.\ 1992).  It is also a known HI radio
source: AGC~258125 (Haynes et al.\ 2011)\footnote{The ``Arecibo General
  Catalog'' is a private database maintained by Martha P.\ Haynes and Riccardo
  Giovanelli.} from the Arecibo Legacy Fast Arecibo L-band Feed Array
(ALFALFA) survey.   
LEDA~87300 has a reported heliocentric velocity of 7278 km s$^{-1}$ in
Schombert et al.\ (1992), which was confirmed by Haynes et al.\ (2011) who
report a value of 7283$\pm$71 km s$^{-1}$.  Correcting for (Virgo +
Great-Attractor + Shapley)-infall (Mould et al.\ 2000, via
NED\footnote{http://ned.ipac.caltech.edu/}), this heliocentric velocity
equates to a recessional velocity of 7936$\pm$77 km s$^{-1}$, or a Hubble
expansion redshift of 0.02647.  Using $H_0 = 70$ km s$^{-1}$ Mpc$^{-1}$,
$\Omega_m = 0.3$ and $\Omega_{\Lambda} = 0.7$, this gives a (comoving radial)
distance of 112.7 Mpc and a physical scale of 532 parsec per arcsecond (Wright
2006).

\section{Optical Image Analysis}\label{Sec_Data}

\subsection{Data}

\begin{figure}
\includegraphics[width=\columnwidth]{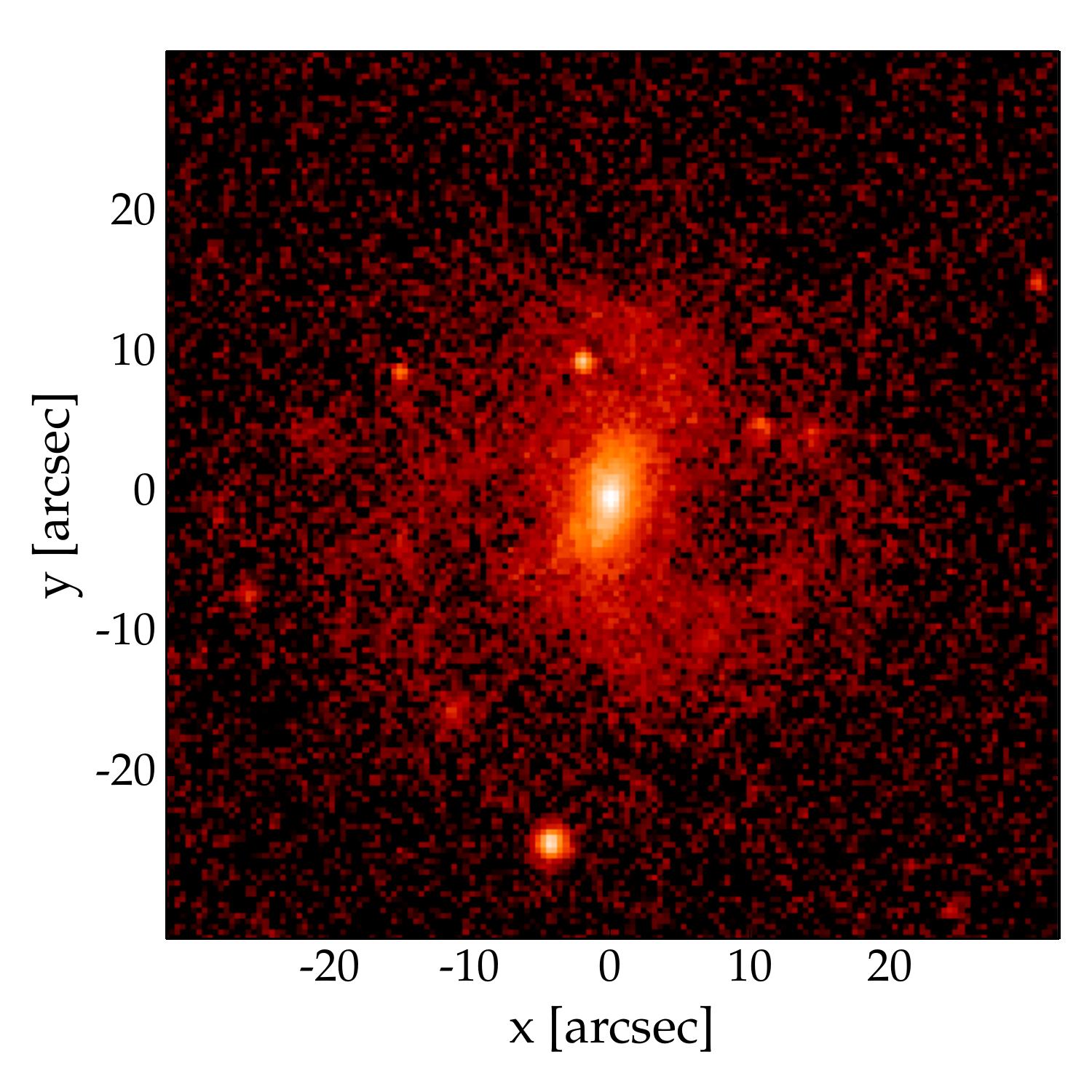}
\caption{Stacked {\it SDSS} $g^{\prime}$-, $r^{\prime}$-, and $i^{\prime}$-band
  image of LEDA~87300 (SDSS J152304.96+114553.6) to increase the $S/N$ ratio
  and better reveal the barred spiral galaxy.  This image guided the 
  galaxy decompositions subsequently performed on the $g^{\prime}$- 
  and $r^{\prime}$-images.  The scale is such that 
  $10\arcsec = 5.32$ kpc in this false color image. East is up and North is to the right. 
}
\label{Fig2}
\end{figure}

While there is a 2008 Canada France Hawaii Telescope (CFHT)/MegaCam image of
the field containing LEDA~87300 (Yee et al.\ Proposal ID 07AC22),
unfortunately LEDA~87300 falls right in the middle of the CCD mosaic gap
(which we mention to spare some readers from looking for it there).  We 
have therefore obtained images available from the Sloan Digital Sky Survey
({\it SDSS}) Data Release Nine (Ahn et
al.\ 2012)\footnote{http://www.sdss3.org/dr9}.  Figure~\ref{Fig2} displays an
image of LEDA 087300 created by us using data from two Sloan runs 
in the DR9 Science Archive\footnote{http://dr9.sdss3.org/fields/},
identified by the Run, Camcol, and Field numbers (005322, 2, 0142) and
(003996, 6, 0175).

\begin{figure}
\includegraphics[width=\columnwidth]{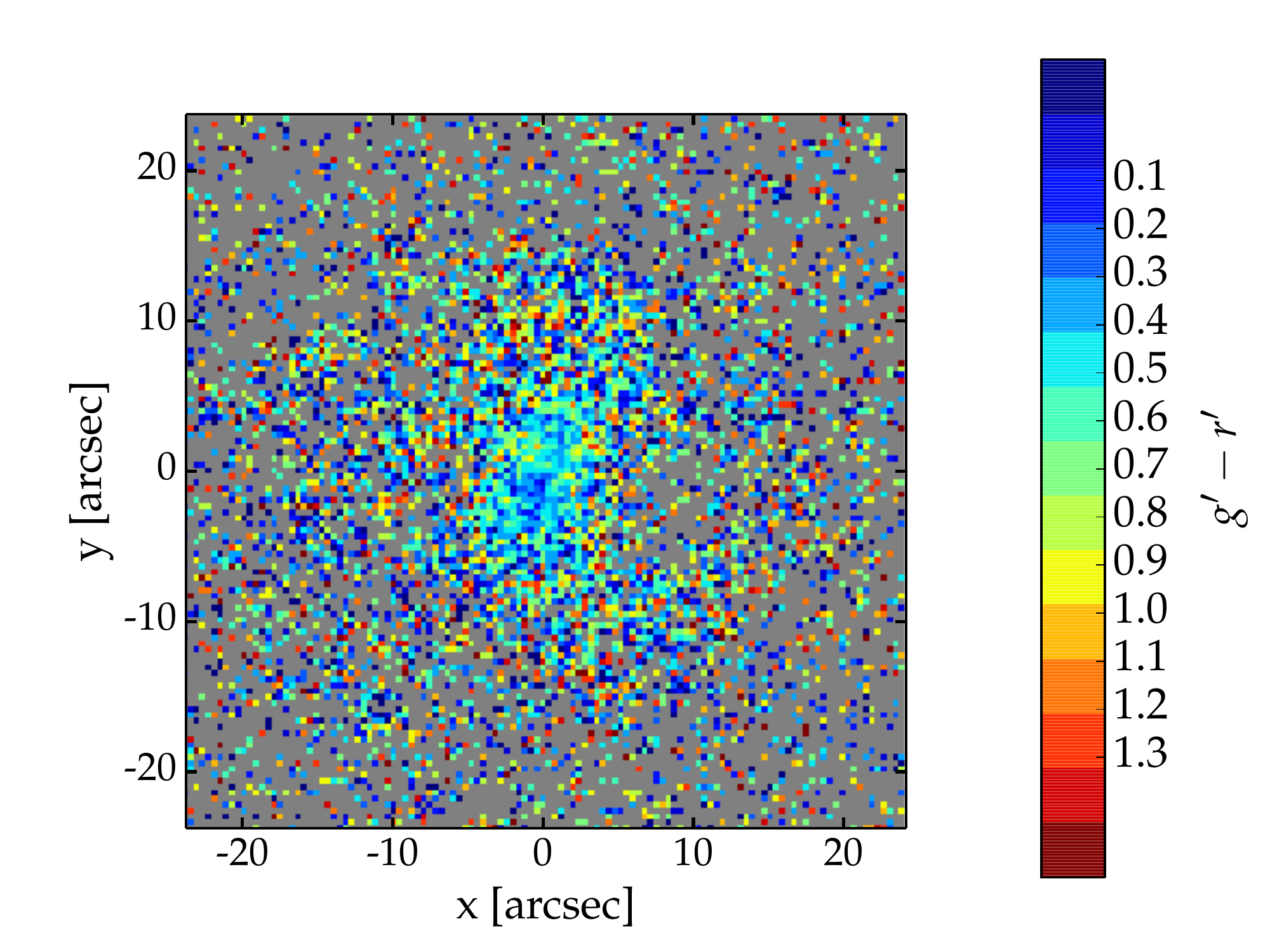} 
\caption{Color map of LEDA~87300, revealing a 
  relatively red bulge ($g^{\prime} - r^{\prime} \sim$ 0.5) 
  and blue bar ($g^{\prime} - r^{\prime} \sim$ 0.1--0.2).  The data at larger
  radii is noisy but has been kept to aid with the recognition of the spiral arms. 
  Correcting for Galactic dust extinction makes the entire image bluer by 0.038. 
  East is up and North is to the right. 
}
\label{Fig3}
\end{figure}

We aligned the $g^{\prime}$-, $r^{\prime}$-, and $i^{\prime}$-band images from
these two runs by rotating and shifting the images, and then we stacked the
images from all three filters to increase the signal-to-noise.  It is apparent
from Figure~\ref{Fig2} that LEDA~87300 is a barred spiral galaxy.  The
morphology of LEDA~87300 is not unusual, and there are plenty of other nearby
($D < 30$ Mpc), relatively face-on, late-type barred spiral galaxies which
resemble LEDA~87300, such as NGC~3319, NGC~4519, NGC~7741 and UGC~6983 (Tully
1988).  In Figure~\ref{Fig3} we show the $g^{\prime} - r^{\prime}$ color map
for LEDA~87300.  This was created after convolving the images with the better
`seeing' with a Gaussian function in order to degrade them to the spatial
resolution (1$\arcsec$.27) of the poorest image.  While the color map is noisy
in the outer regions of this pixel-by-pixel representation, without binning,
the inner regions already hint at a somewhat blue bar upon which a small,
relatively redder spheroid may be superimposed.
 
\begin{figure*}
\begin{center}
\includegraphics[width=0.33\textwidth]{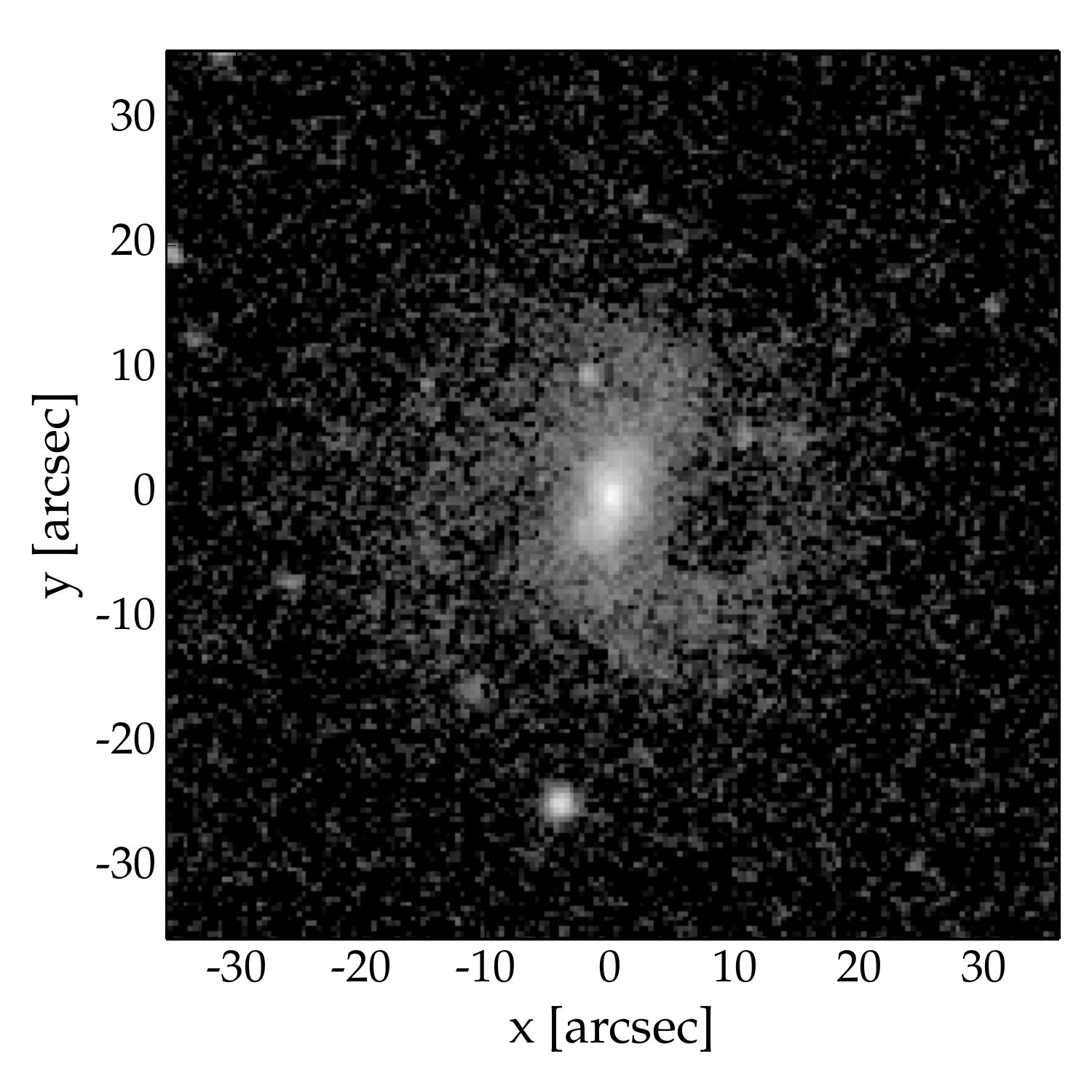}
\includegraphics[width=0.33\textwidth]{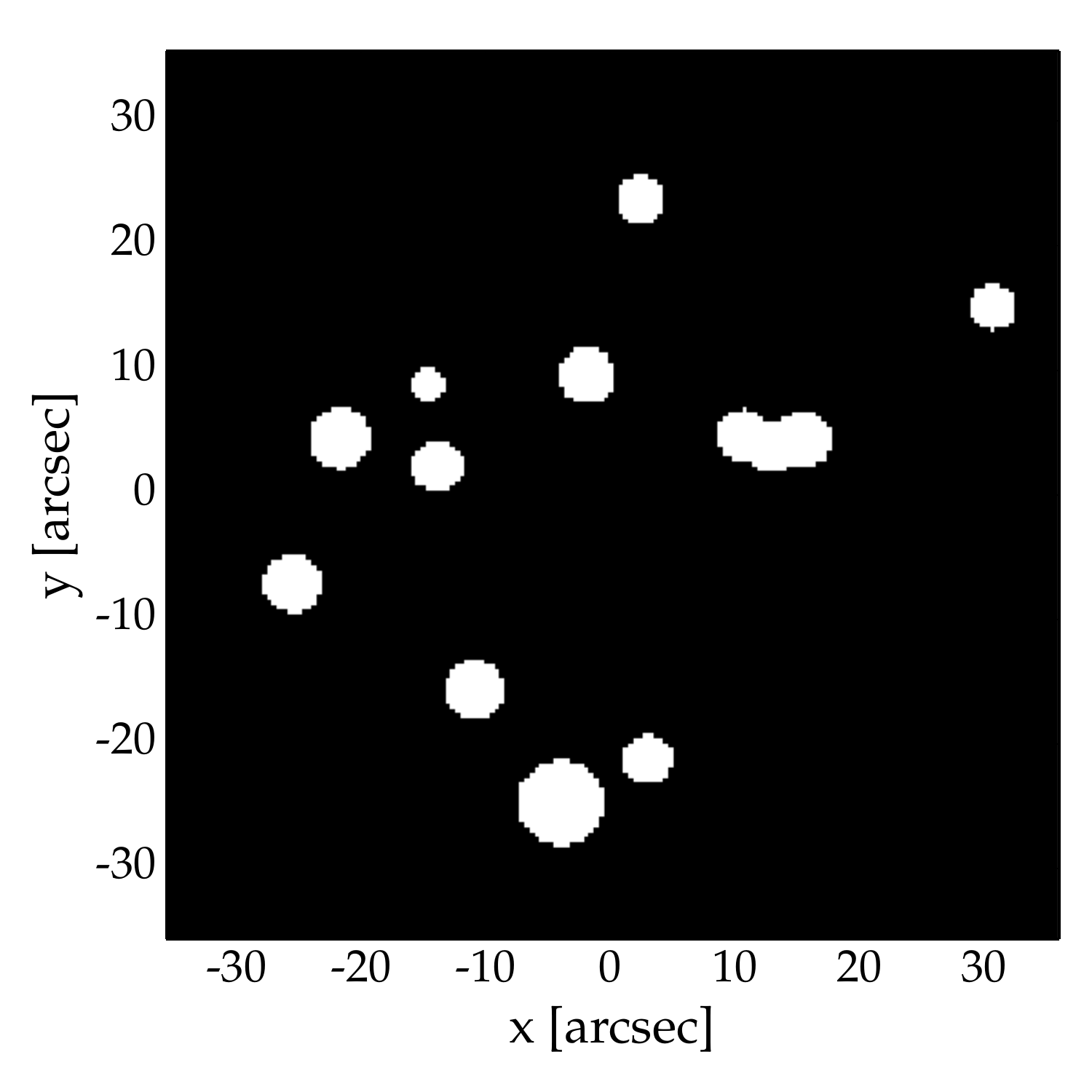}
\includegraphics[width=0.33\textwidth]{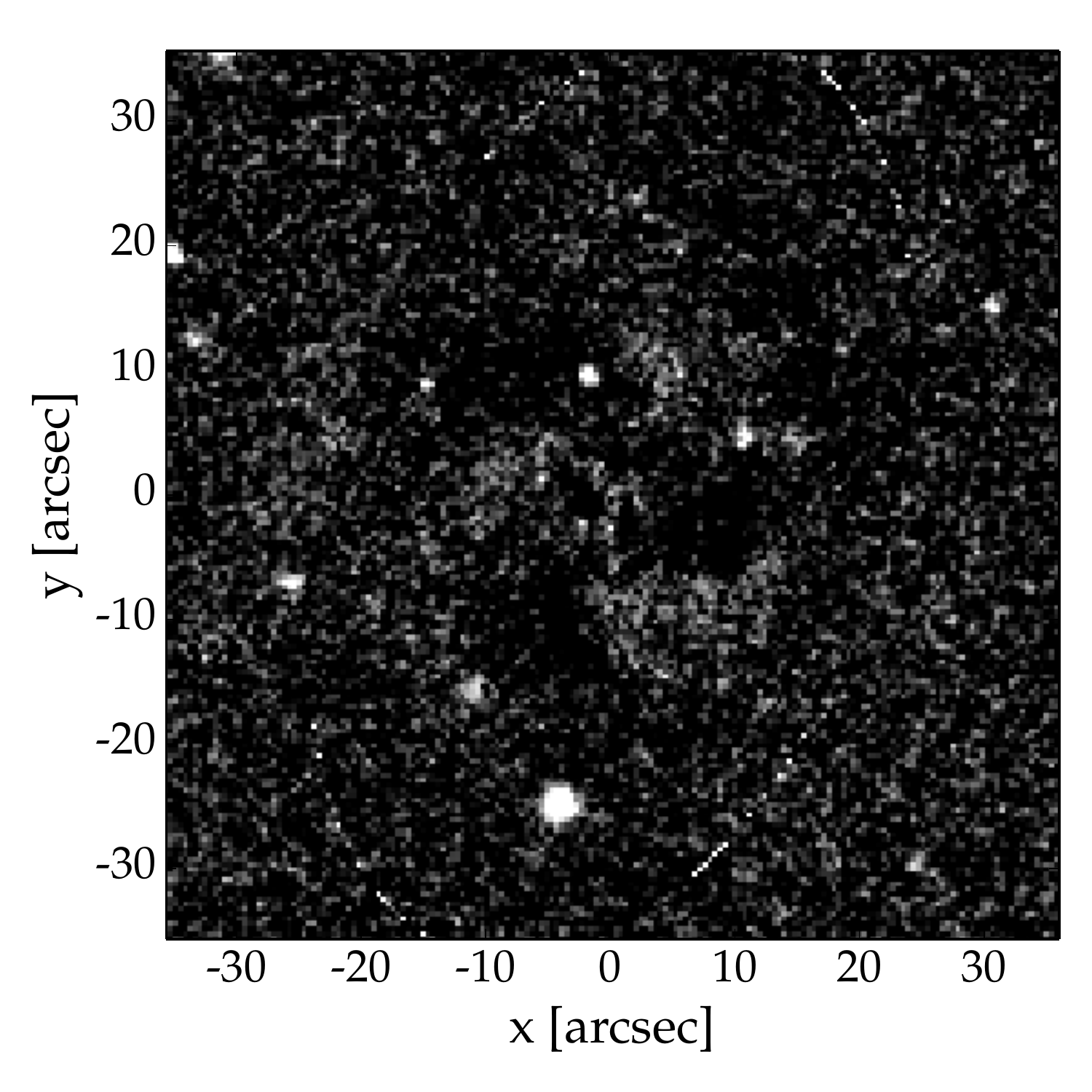} 
\caption{
Left panel: {\it SDSS} $r^{\prime}$-band image of LEDA~87300, with a logarithmic flux scaling. 
Middle panel: Mask used when extracting, and modelling, the galaxy light. 
Right panel: Residual image after subtracting the 
galaxy as modelled with the new {\tt cmodel} (construct model) 
and {\tt Isofit} tasks in IRAF, rather than using the 
{\tt bmodel} (build model) and {\tt ellipse} tasks 
(Jedrzejewski 1987, see Ciambur 2015 for details). 
The display stretch has been adjusted here and is now linear, to better show the
residuals at all radii. 
A three-armed spiral structure is apparent.  
The scale is such that $10\arcsec = 5.32$ kpc, 
} 
\label{Fig4}
\end{center}
\end{figure*}

For our analysis, we used the 
two available datasets to generate a single image for each of the two
bands that we modelled. We did this for the $g^{\prime}$- and
$r^{\prime}$-band images, but not the $i^{\prime}$-band image because it is
too faint.  
Galaxy images are not always well approximated by 2-dimensional models whose 
components have fixed ellipticity ($\epsilon$) and position angle (P.A.).  
Instead, we have used 1-dimensional `light profiles' and the associated 
geometrical profiles which track a galaxy's variation in $\epsilon$, P.A.,
and isophotal deviation from ellipses --- as quantified using Fourier
harmonics.  Collectively, this suite of 1-dimensional profiles describe and recreate
the symmetrical elements of a 2D image.  
The {\tt Isofit} task (Ciambur 2015) was used to obtain both
the major-axis and the geometric-mean-axis (equivalent to a circularized)
light profile, and its accompaniments at each radius (i.e.\ position angle,
ellipticity, Fourier terms).  As can be seen in the right hand panel of
Figure~\ref{Fig4}, the 1D profiles perform well at representing the galaxy and
reveal three spiral arms in the residual map.  To avoid redundancy, we do not
show the similar set of $g^{\prime}$-band image, mask and residual map,
although we do note that the $g^{\prime}$-band profiles could only be reliably 
extracted to half the radial extent of the $r^{\prime}$-band profile.

\subsection{Modelling}

Having extracted the $g^{\prime}$- and $r^{\prime}$-band light profiles (shown in
Figures~\ref{Fig5} and \ref{Fig6}), they have then been decomposed into model
components (point-source, bar, bulge, disk) using the {\it Profiler} software
(Ciambur 2016, in preparation).  `Profiler' finds the optimal decomposition of
a light profile through an iterative minimization process using the
Levenberg-Marquardt algorithm (Marquardt 1963).  At each iteration step, the
components are summed together and then convolved with the point spread
function (PSF) --- using a hybrid (Fast Fourier Transform)-based and numerical
convolution scheme --- to generate a model profile.  Profiler can use a Gaussian,
Moffat, or any user-supplied vector to describe the PSF.  
The best-fit values for the galaxy model parameters are then found by
minimizing the root mean square scatter $\Delta_{rms} = \sqrt{ \sum_{i=1,N}
  ({\rm data}_i-{\rm model}_i)^2/(N-\nu)}$, where $N$ is the number of fitted
data points in the light profile and $\nu$ is the number of model parameters.

It should be remembered that $\Delta_{rms}$ reflects the quality of the global
fit.  It can therefore be beneficial to also watch the behavior of the
residual profile over specific regions of interest.  While a
(signal-to-noise)-weighted fitting scheme can be used, it is generally
preferable not to do so due to the myriad of potential biases (above the
Poissonian noise expected from the galaxy) 
affecting the inner light profiles of galaxies where the signal is highest.
Sources of bias 
can include central dust, a faint AGN, additional components such
as a nuclear star cluster or a nuclear disk, and errors 
in the PSF, all of which become more important when using a 
(signal-to-noise)-weighted scheme. 
While this danger is considerably reduced through the use $\Delta_{rms}$, one
is more susceptible to biases from an incorrect sky-background subtraction
when using this statistic.  In the case of LEDA~87300, it has a
non-exponential disk which we avoided by restricting the radial range used,
and thus we are not using data that could be overly susceptible to errors from
the sky-background subtraction process.  In passing we note that even if we
had modelled the disk with a broken, inner and outer, exponential function
(e.g., Pohlen et al.\ 2002; Pohlen \& Trujillo 2005), it would not have
affected our results for the spheroid.

The {\tt imexamine} task in IRAF was used to fit 
Gaussian profiles to a dozen stars in each of the final $g^{\prime}$- and
$r^{\prime}$-band images, and the mean Full Width Half Maximum (FWHM) used to
characterize our Gaussian PSF.  Initially we fit the same model components as
BRGG15, namely a point-source, an exponential disk, plus a single S\'ersic
(1963) $R^{1/n}$ function\footnote{
A review of the S\'ersic function, which was popularized
by Caon et al.\ (1993) and reproduces an exponential and Gaussian profile when
$n$ equals 1.0 and 0.5, respectively, can be found in Graham \& Driver
(2005).} for the combined bar+bulge (``barge''), as shown in Figure~\ref{Fig5}.  

We obtained a notably smaller $r^{\prime}$-band S\'ersic index for the barge
than BRGG15 (0.41 compared to 1.13), and a barge luminosity which is less than
two times as bright.  Our uncorrected $r^{\prime}$-band barge magnitude of
19.29 mag and disk magnitude of 16.82 mag give a barge-to-disk flux ratio of
$\sim$0.10.  Given that the bulge-to-disk flux ratio is typically just a few
percent for Scd--Sm galaxies (e.g., Graham \& Worley 2008), this is additional
evidence (beyond simply looking at the 2D image and the 1D profile) that the
bar is likely contributing to the barge light\footnote{We obtained a disk
  $g^{\prime}-r^{\prime}$ color of 0.41$\pm$0.03 and a bluer barge color of
  0.37, albeit with an uncertainty of 0.05.}.  While at some stage bars become
bulges within the regime of secular disk evolution (e.g., Hohl 1975; Combes \&
Sanders 1981; Combes et al.\ 1990), and thus one could perhaps argue for the
use of barge light rather than bulge light, in the following section 
we have proceeded by performing a 
bulge$+$bar$+$disk decomposition, as done for the galaxies with supermassive black
holes (e.g., L\"asker et al.\ 2014a; Savorgnan \& Graham 2015a, and references
therein).

\subsubsection{A bar$+$bulge$+$disk decomposition}\label{Sec_bbd}

We used a modified Ferrers profile to describe the bar (Ferrers 1877; Peng et
al.\ 2010).  The best-fitting bulge$+$bar$+$disk model parameters are provided
in Table~1.  Of immediate note is the improved quality of the fit, evinced by
the smaller residuals around the inner $\sim$5$\arcsec$ and the reduced rms 
scatter (Figure~\ref{Fig6}).  With a major-axis, $r^{\prime}$-band half light radius equal to
two-thirds of a kpc, the central component is neither an AGN point source nor
a star cluster but instead the bulge component of this galaxy.  Its size is
typical for late-type spiral galaxies (e.g., Graham \& Worley 2008), further
suggesting that there is nothing unusual with LEDA~87300.  

\begin{figure}
\begin{center}
\includegraphics[angle=0, width=\columnwidth]{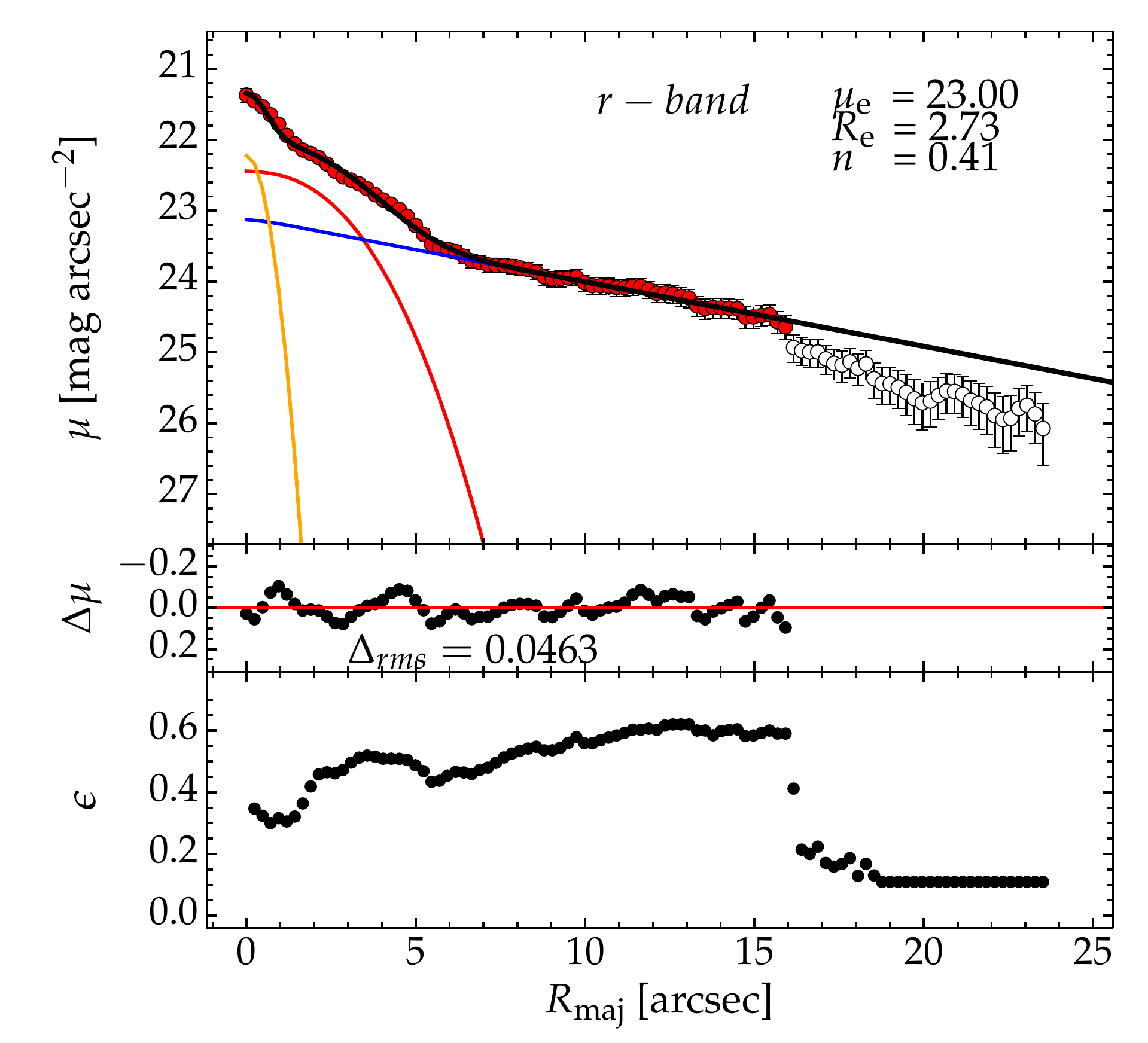}
\caption{The major-axis $r^{\prime}$-band light profile taken from the left
  panel of 
  Figure~\ref{Fig4} is modelled here with a 3-parameter S\'ersic function
  (red) for the ``barge'' (bar+bulge), a 2-parameter exponential for the disk
  (blue), plus a 2-parameter point-source for a necessarily bright AGN 
  (orange).  The best-fitting S\'ersic parameters are inset in the figure, and 
  the error bars on the data points show our 3$\sigma$ sky-background
  uncertainty (see
  section~\ref{Sec_bbd} for details).  Rather than fit a broken exponential
  model for the disk (e.g., Pohlen \& Trujillo 2005), we truncate the outer
  data and fit for just the inner exponential.  The lower panel shows the 
  ellipticity profile. } 
\label{Fig5}
\end{center}
\end{figure}

\begin{figure*}
\begin{center}
\includegraphics[angle=0, width=0.33\textwidth]{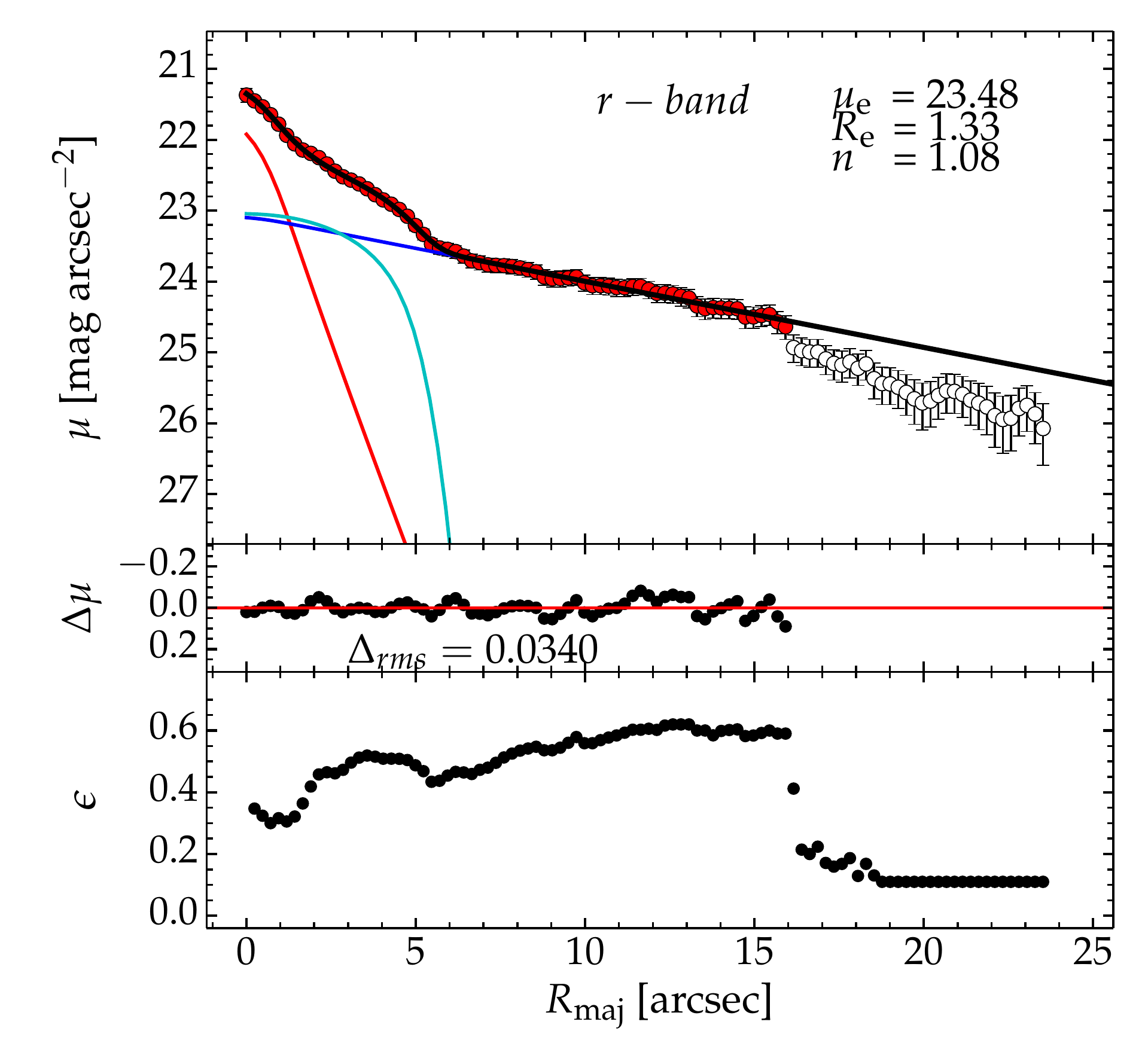}
\includegraphics[angle=0, width=0.33\textwidth]{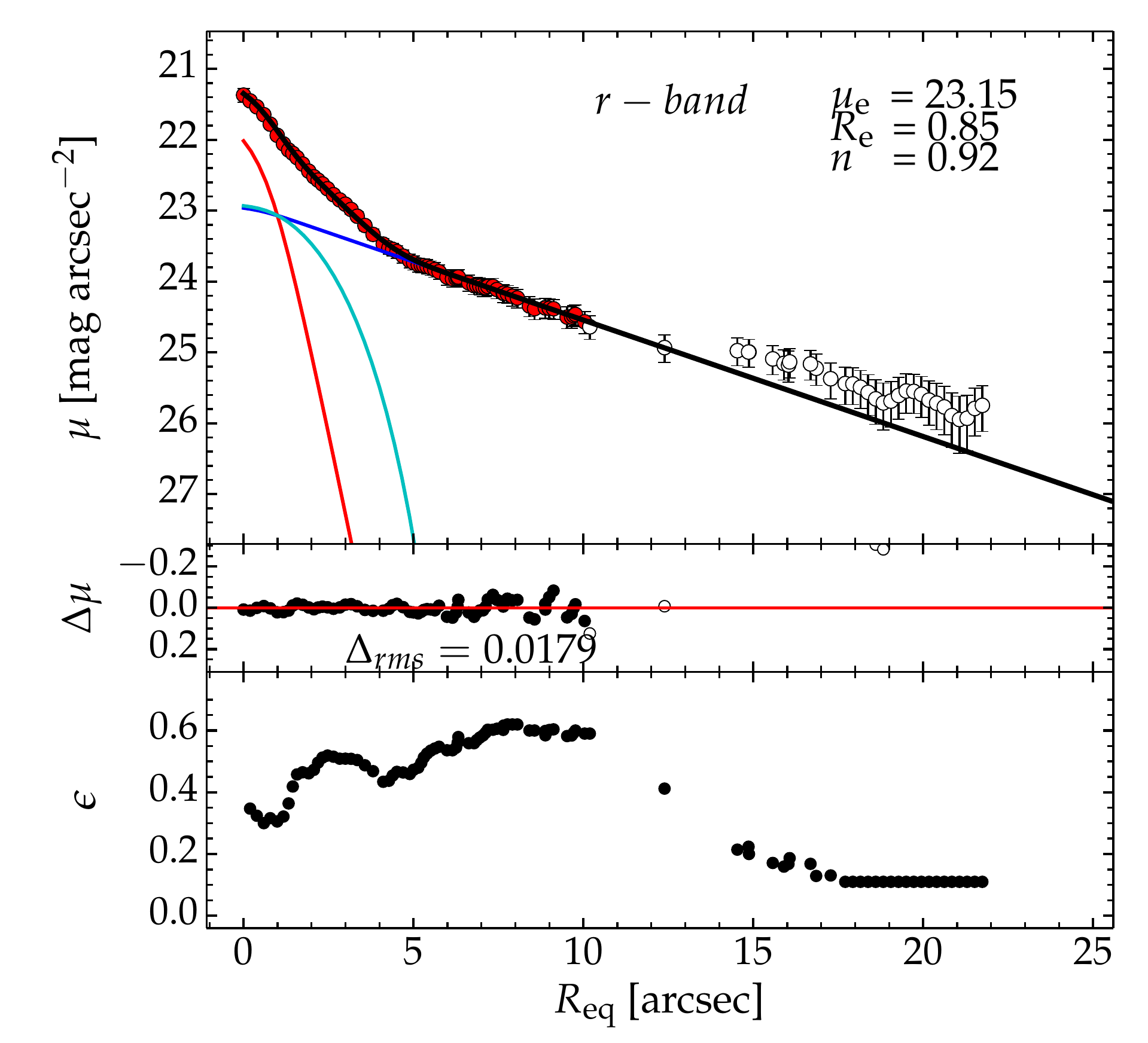}
\includegraphics[angle=0, width=0.33\textwidth]{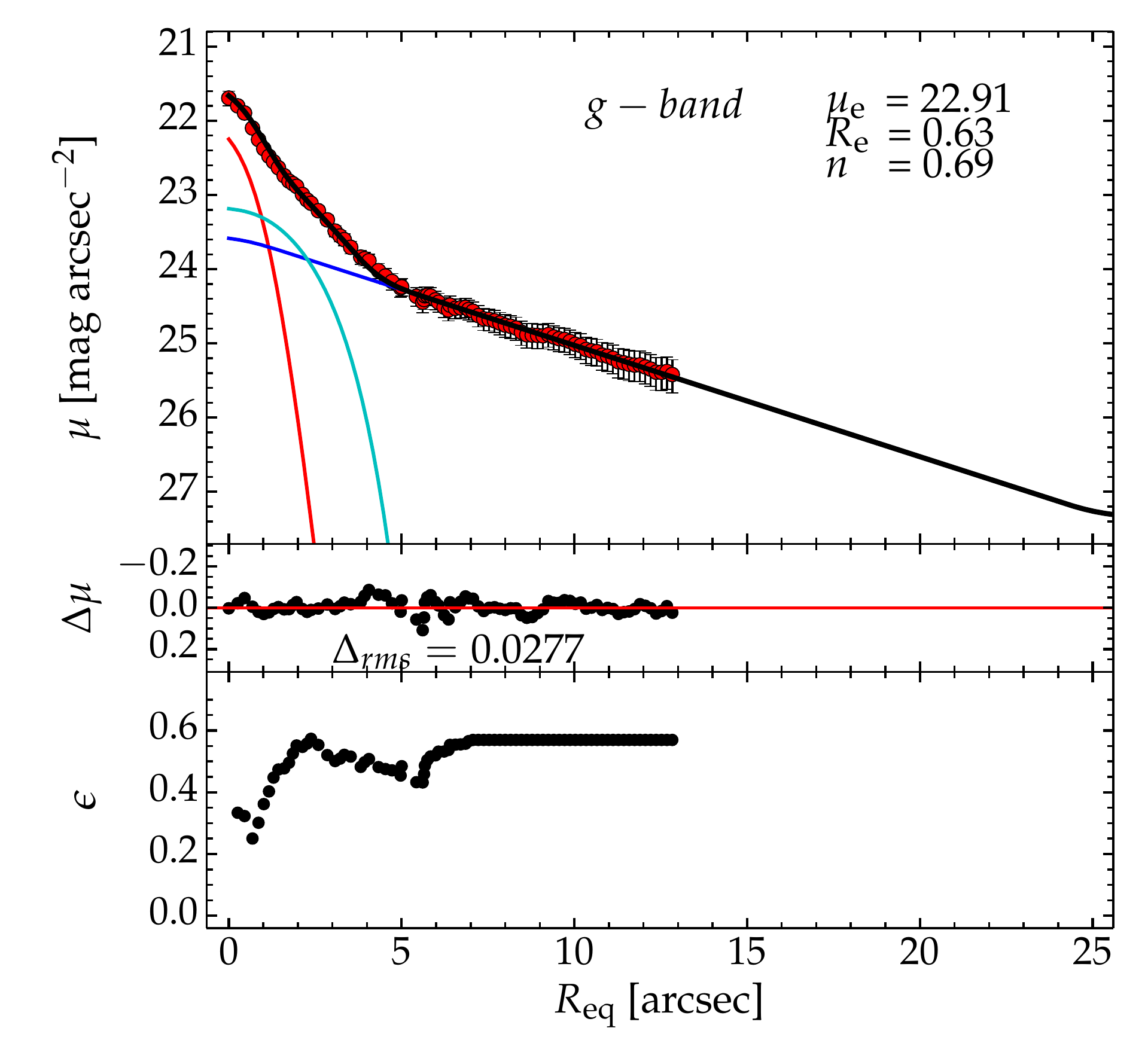} 
\caption{
Left panel: Major-axis $r^{\prime}$-band light profile 
taken from the left panel of Figure~\ref{Fig4} is modelled
with a S\'ersic function for the galaxy's spheroidal component (red), an
exponential function for the disk (dark blue), and a Ferrers function for the
bar (cyan). 
While the bulge parameters (effective surface brightness, effective half light
radius, and S\'ersic index $n$) are inset in the figure to aid comparison, the
parameters of all the model components are listed in Table~1. 
The residual profile is shown in the panel immediately below, and the 
ellipticity profile derived using the new IRAF task {\tt Isofit} (Ciambur
2015) is displayed at the bottom. 
Middle panel: Geometrical mean ($\sqrt{ab}$) `equivalent axis', $r^{\prime}$-band
light profile. 
Right panel: Same as the middle panel but for the $g^{\prime}$-band light profile. 
Note:  The plotted models have been convolved with the PSF, which explains why the 
(deconvolved) $\mu_{\rm e}$ and $R_{\rm e}$ of the bulge {\it appear} not to
match the PSF-convolved model. 
The first peak in the ellipticity
profile shows the characteristic rising dominance of a bar as it extends
beyond a rounder central bulge, while the second peak pertains
to the spiral arms in the disk. The radial scale is such that 1$\arcsec=532$ pc. 
}
\label{Fig6}
\end{center}
\end{figure*}

BRGG15 included a
central point source in their model (see their Figure~3 and our
Figure~\ref{Fig5}), but we do not see evidence of a possible fourth
component after our new model has described the three major 
components of this galaxy.  Future {\it Hubble Space Telescope} images would
be valuable for assessing the contribution/existence of any optical AGN point
source.  A nuclear star cluster may also be expected, and the prescription in
Graham (2015b) suggests that it would be a few million solar masses if 
associated with the black hole in LEDA~87300, and thus it would have an
$r^{\prime}$-band magnitude prior to dust extinction of $\approx -24$ mag
(assuming a stellar mass-to-light ratio of unity).  This is too faint for us
to detect in the {\it SDSS} image.  A nuclear disk is yet another option, but
again, our attempts to fit a fourth component did not detect anything.

\setcounter{table}{0}

\begin{table*}[ht]
\label{Tab_Comp}
\centering
\caption{$r^{\prime}$-band galaxy model parameters.}
\begin{tabular}{lcccccccccc} 
\hline\hline
Component & $\mu_{\rm e}$ or $\mu_0$ & $R_{\rm e}$   & $h_{\rm disk}$ or $R_{0,bar}$ & $(\alpha, \beta)_{bar}$ & $n_{\rm Sersic}$ & $m$    & $m^{corr1}$ & Mag$^{corr1}$ & $m^{corr2}$ & Mag$^{corr2}$ \\
         & [mag arcsec$^{-2}$]    &[arcsec / kpc] & [arcsec / kpc]            &                        &               & [mag]  &  [mag]    &  [mag]      &    [mag]   &  [mag]      \\
 (1)     &   (2)                 &  (3)          & (4)                       &      (5)               &  (6)          & (7)    &  (8)      &   (9)       &    (10)    &  (11)        \\
\hline  

\multicolumn{9}{c}{{\it SDSS} $r^{\prime}$-band major-axis} \\
Spheroid &  23.48                &  1.33 / 0.71  &  ...                      &        ...             & 1.08          & ...    &  ...      &  ...        &   ...     &  ...      \\
Disk     &  23.06                &  ...          &  11.57 / 6.16             &        ...             & 1.00          & ...    &  ...      &  ...        &   ...     &  ...      \\
Bar      &  23.05                &  ...          &  5.32 / 2.83              &      (1.94, 0.01)      &  ...          & ...    &  ...      &  ...        &   ...     &  ...      \\
\multicolumn{9}{c}{{\it SDSS} $r^{\prime}$-band equivalent-axis} \\
Spheroid &  23.15                &  0.85 / 0.45  &  ...                      &        ...             & 0.92          & 20.85  &  20.71    &  $-14.55$   & 20.05     &  $-15.21$  \\
Disk     &  22.91                &  ...          &  6.49 / 3.45              &        ...             & 1.00          & 16.85  &  16.71    &  $-18.55$   & 16.54     &  $-18.72$  \\
Bar      &  22.90                &  ...          &  5.82 / 3.10              &      (9.99, 0.01)      &  ...          & 20.44  &  20.30    &  $-14.96$   & 20.13     &  $-15.13$  \\
Galaxy   &  ...                  &  ...          &  ...                      &        ...             &  ...          & 16.78  &  16.64    &  $-18.62$   & 16.46     &  $-18.80$  \\
\multicolumn{9}{c}{{\it SDSS} $g^{\prime}$-band equivalent-axis} \\
Spheroid &  22.91                &  0.63 / 0.34 &  ...                       &        ...             &  0.69         & 21.39  &  21.21    &  $-14.05$   & 20.35     &  $-14.91$  \\
Disk     &  23.53                &  ...          &  7.23 / 3.85              &        ...             &  1.00         & 17.24  &  17.06    &  $-18.20$   & 16.84     &  $-18.42$  \\
Bar      &  23.15                &  ...          &  4.61 / 2.45              &      (5.79, 0.01)      &   ...         & 20.68  &  20.50    &  $-14.76$   & 20.28     &  $-14.98$  \\
Galaxy   &  ...                  &  ...          &  ...                      &        ...             &   ...         & 17.17  &  16.99    &  $-18.27$   & 16.75     &  $-18.51$   \\
\hline 
\end{tabular}

Column~1: Model components fit to the galaxy light-profile. 
Column~2: Effective surface brightness of the spheroid, or central surface brightness of the disk or bar. 
Column~3: Effective half light radius. 
Column~4: Disk or bar scale-length. 
Column~5: Shape parameters of the Ferrers profile. 
Column~6: S\'ersic index. 
Column~7: Total observed, i.e.\ uncorrected, apparent magnitude (AB mag). 
Column~8: Galactic-extinction and redshift-dimming corrected apparent magnitude.
Column~9: Absolute magnitude associated with column~8. 
Column~10: Apparent magnitude from column~8 now further corrected for dust internal to the galaxy and the galaxy's 
inclination (Driver et al.\ 2008).  
Column~11: Fully corrected absolute magnitude associated with column~10. 
The magnitudes correspond to the fitted components integrated to infinity. 
\end{table*}

Integrating the surface brightness profiles (to $R=\infty$) of the individual
model components that were fit to the equivalent-axis light profile
(Figure~\ref{Fig6}) has the advantage that it can be done using spherical
symmetry (see the Appendix in Ciambur 2015 to understand how this recovers
the total galaxy light).  The resultant luminosity of each component is
included in Table~1, and collectively gave a total, uncorrected
$g^{\prime}$-band magnitude of 17.17 mag for the galaxy.  For reference,
Garnier et al.\ (1996) reported $B_T= 17.05$ mag (RC3 system, de Vaucouleurs
et al.\ 1991) while the `SDSS Model' magnitude from Data Release 6
(Adelman-McCarthy et al.\ 2008)\footnote{http://classic.sdss.org/dr6} gave a
$g^{\prime}$-band magnitude of 17.08 mag (AB).  Our preferred band is of
course the redder $r^{\prime}$-band, however the $g^{\prime}$-band parameters
derived here are used to a) check for consistency with the literature as just
done and b) obtain a $g^{\prime} - r^{\prime}$ color for the bulge (see later
in this section).

We explored the impact on the $r^{\prime}$-band bulge magnitude due to the
uncertainty in the sky-background.  
Due to the large {\it SDSS} images, and thus the large number of $N$ sky pixels
surrounding LEDA~87300, the uncertainty in the median sky value, equal to the
standard deviation of the $N$ sky values ($\sigma_{\rm sky}$) 
divided by $\sqrt{N}$, is tiny.  As 
such, it is the variation and uncertainty of the individual sky values in the pixels
co-occupied by the galaxy which are important.  Having $n$ pixels
make up each isophote reduces this uncertainty to $\sigma_{\rm sky}/n$ at each
isophotal radius.  Nonetheless, 
at large radii, this error in the sky-subtraction (such that the median sky-value for
the isophote does not match the global median sky-value that was subtracted) 
dominates over the Poissonian scatter in the galaxy light. 
We repeated the galaxy decomposition with the uncertainty ($\sigma_{\rm sky}/n$) 
added to, and then subtracted from, the optimal light profile.  
Given that our fit was performed over the 
inner 10--16 arcseconds, this did not have a big impact, with the 
spheroid's magnitude changing by ($-0.16, +0.12$) mag. 
Adding and subtracting three times the uncertainty in the sky-background, 
the bulge magnitude changed by ($-0.37, +0.31$) mag, the bar 
magnitude changed by just ($+0.05, -0.05$) mag, and the disk magnitude 
changed by ($-0.24, +0.15$) mag. 
 
Unsurprisingly, changing the width of the PSF by a rather large $\pm$25\% had 
almost no affect on the disk parameters, with the magnitude changing by just
$\pm$0.03 mag.  While the bar was found to be similarly stable, the spheroid
magnitude changed by $\pm$0.4 mag ($\pm$45\%) and the S\'ersic index changed
by $\pm$6\%.

\begin{table*}[ht]
\centering
\label{Tab_mass}
\caption{Galaxy model's component masses.}
\begin{tabular}{lcccc|cccc}
\hline\hline
Component & $L^{corr1}_{r^{\prime}}$ & $(g^{\prime}-r^{\prime})^{corr1}$ & $(M_*/L_{r^{\prime}})^{corr1}$ &  $M^{corr1}_*$        &  $L^{corr2}_{r^{\prime}}$  & $(g^{\prime}-r^{\prime})^{corr2}$ & $(M_*/L_{r^{\prime}})^{corr2}$ & $M^{corr2}_*$       \\
          & [$10^{8}\rm ~L_{\odot}$] &                          & [$M_{\odot}/L_{\odot}$]      & [$10^{8} ~M_{\odot}$] &  [$10^{8}\rm ~L_{\odot}$] &                           & [$M_{\odot}/L_{\odot}$]     & [$10^{8} ~M_{\odot}$] \\ 
 (1)      &   (2)              &       (3)                    &    (4)                   &    (5)             &        (6)              &         (7)               &    (8)                   &    (9)            \\
\hline
Spheroid  &   0.48             &  0.50                        &   1.76                   &    0.84            &        0.88             &    0.30                   &     1.06                 &   0.93            \\
Disk      &  19.12             &  0.35                        &   1.20                   &   23.00            &       22.20             &    0.29                   &     1.03                 &   22.92            \\
Bar       &   0.70             &  0.20                        &   0.82                   &    0.58            &        0.81             &    0.14                   &     0.71                 &   0.57            \\
\hline
\end{tabular}

Column~1: Model components fit to the galaxy light profile.
Column~2: $r^{\prime}$-band solar luminosity from column~8 of Table~1. 
Column~3: $g^{\prime}-r^{\prime}$ color derived from column~8 of Table~1. 
Column~4: Adopted stellar mass-to-light ratio following Bell et al.\ (2003). 
Column~5: Stellar mass.
Column~6: $r^{\prime}$-band solar luminosity from column~10 of Table~1.
Column~7: $g^{\prime}-r^{\prime}$ color derived from column~10 of Table~1.
Column~8: Adopted stellar mass-to-light ratio following Bell et al.\ (2003). 
Column~9: Stellar mass. 
\end{table*}

We next corrected the observed, apparent magnitude for the usual influences.
First, in the {\it SDSS} $g^{\prime}$- and $r^{\prime}$-band there is 0.122
and 0.084 mag, respectively, of extinction in the direction of LEDA~87300 due
to dust in our Galaxy (Schlafly \& Finkbeiner 2011).  We also brightened the
observed magnitudes by $5\log(1+z) = 0.06$ mag, allowing for cosmological
redshift dimming.  Together, this gives us the first corrected magnitudes
shown in Table~1.  These entries can be compared with the results from BRGG15.
In particular, their stated ``bulge'' (barge) luminosity is 7.7 times brighter
than our bulge luminosity, and more than 3 times as bright as our combined
bulge plus bar luminosity.  This is because of the way the bar in BRGG15's
modelling resulted in a S\'ersic component which extended beyond the end of
the bar, eating into the disk light.

With the corrections mentioned above, we obtain a $g^{\prime} - r^{\prime}$ 
disk color of 0.35, consistent with the value of $0.41\pm0.07$ 
reported in BRGG15.  We additionally obtain a bulge color of $0.50\pm0.2$, 
and a blue bar colour of $0.20\pm0.2$. 
Unlike the long bars of early-type spiral galaxies 
with bar colors that are similar to or redder than their host (e.g., Prieto et
al.\ 2001), 
the shorter bars of late-type spiral galaxies tend to be bluer than their galaxy
due to recent and ongoing star formation along the bar, revealing that 
they are young and growing (e.g., 
Elmegreen \& Elmegreen 1985; Kennicutt 1994; Phillips 1996; Gadotti \& de
Souza 2006; Fanali et al.\ 2015, and references therein).  
The sites along the 
bar in LEDA~87300 with a $g^{\prime} - r^{\prime}$ color of $\sim$0.1--0.2 (Figure~\ref{Fig3})
are indicative of this. 

The second set of corrections that we apply are the combined
inclination and dust corrections from Driver et al.\ (2008).  
At $R_{\rm maj}\approx 16\arcsec$, 
beyond the radial range where the spiral arms dominate the ellipticity of
LEDA~87300, 
the ellipticity profile rapidly drops to about 0.1$\pm$0.1 by 20$\arcsec$, although
the low ratio of signal-to-noise makes it hard to know the exact value. 
For a
disk inclination corresponding to an isophotal ellipticity of 0.1 (i.e.\ close
to face-on), the bulge needs to be corrected / brightened by 0.86 and 0.66 mag
in the $g^{\prime}$- and $r^{\prime}$-band, while both the disk and bar (made
out of the disk) should be 
brightened by 0.22 and 0.17 mag in the $g^{\prime}$- and $r^{\prime}$-band,
respectively\footnote{If the adopted ellipticity is wrong by 0.1, then the corrections to the bulge
magnitude will change by just 0.02--0.06 mag; although it should be noted that
while the mean corrective prescription from Driver et al.\ (2008) is appropriate for an ensemble of
galaxies, its applicability for individual galaxies will vary.  This makes it
difficult to assess the error on the bulge magnitude due to dust.}. 
The second last column in Table~1 shows the
apparent magnitude of the galaxy's components after this additional
correction is applied.  
These corrected apparent magnitudes were then
converted into absolute magnitudes using a distance modulus of 35.26, 
and used to obtain a second set of colors. 
The $r^{\prime}$-band absolute magnitudes were converted into solar
luminosities using an absolute magnitude for the Sun of 
$M_{r^{\prime},\odot} = 4.65$ AB mag (Table~2).

\subsubsection{Stellar Masses}

We use the $g^{\prime}-r^{\prime}$ color to estimate the $r^{\prime}$-band stellar
mass-to-light ($M_*/L_r$) ratios\footnote{We had hoped to use the
$i^{\prime}$-band luminosity and the relation
$\log(M_*/L_i)=-0.68+0.70(g^{\prime}-i^{\prime})$ from Taylor et
al.\ (2011), however, as noted before, the signal-to-noise ratio in the
combined 
$i^{\prime}$-band image was too low.}.  Following BRGG15, we use
the prescription in Bell et al.\ (2003) which is such that
\begin{equation} 
\log_{10}(M_*/L_{r^{\prime}})=1.097(g^{\prime}-r^{\prime})-0.306, 
\label{Eq_Bell}
\end{equation} 
and reportedly has an accuracy of 20\% (Bell et al.\ 2003).  Given that dust
reddening roughly moves one along this relation (see Figure~6 in Bell et
al.\ 2003, and Figure~13 in Driver et al.\ 2007), it can be applied to either
the dust-reddened or the dust-corrected component's luminosity and color.  
The results are shown in
Table~2, using both approaches, and the spheroid masses agree
within $\sim$10\%.  Our preference is to use the dust-corrected luminosities
and masses (column~9 in Table~2).  Using the stellar population
synthesis model of Conroy et al.\ (2009), one has from Roediger \& Courteau
(2015) that $\log(M_*/L_r)=1.497(g^{\prime}-r^{\prime})-0.647$. This yields
mass-to-light ratios for the spheroid which are some 30--40\% smaller than
those obtained via the Bell et al.\ (2003) prescription.  Given the
uncertainties on the bulge luminosity arising from the uncertainty in 
(i) the sky-background, 
(ii) the PSF, 
(iii) the inclination of the disk, 
(iv) the $g^{\prime} - r^{\prime}$ color of the spheroid,
(v) the uncertainty associated with the conversion from light-to-mass, and 
(vi) at some small level even the adopted mask (Figure~\ref{Fig4}), 
we consider the spheroid stellar mass to be accurate to 
within a factor of 2, although it could be as high as 3 if the
inclination/dust correction was terribly wrong.

LEDA~87300 shows up in the GALaxy Evolution eXplorer All-Sky Survey Source
Catalog (GALEXASC J152304.87+114552.0) with
NUV ($\lambda_{\rm eff}=0.227 \mu$m) = 18.69$\pm$0.08 mag.  Brightening for
$\sim$0.20 mag of Galactic extinction and 0.06 mag of cosmological redshift
dimming gives 18.43 mag.  Using the galaxy $r^{\prime}$ magnitude from
column~8 of Table~1, we have that NUV$-r^{\prime} = 1.79$ for the
galaxy.  Together with the galaxy color $g^{\prime}-r^{\prime} = 0.35$,
LEDA~87300 resides in the middle of the `blue cloud' of star forming galaxies
(Chilingarian \& Zolotukhin 2012; their Figure~1), and is more blue than the
average Sc galaxies (Chilingarian \& Zolotukhin 2012; their Figure~3), as
expected for an Sm galaxy.  Its ongoing star formation therefore supports the
low stellar mass-to-light ratios of 0.71--1.76 that we find
(Table~2).

For comparison with BRGG15, and therefore not considering the inclination and internal dust correction, 
we find that BRGG15 obtained a slightly fainter luminosity for 
their disk ($1.4\times 10^9 ~L_{\odot}$) than us ($1.9\times 10^9
~L_{\odot}$: Table~2, column~2), due to their fainter central surface brightness and
smaller scale-length for their fitted disk (see their
Figure~3). However, they also derived a redder disk and higher stellar
mass-to-light ratio than us. This gave them a stellar
mass of $10^{9.3+/-0.1} ~M_{\odot}$ ($=2.0^{+0.5}_{-0.4} \times 10^9
~M_{\odot}$), which agrees well with our optimal disk mass of
$2.3\times 10^9 ~M_{\odot}$. 
As noted before, BRGG15 obtained a notably (7.7$\times$) brighter `spheroid' ($3.7\times 10^8
~L_{\odot}$) than us ($0.48\times 10^8 ~L_{\odot}$), due to their fitted model
being biased by the bar light.  BRGG15 report a ``bulge'' (barge) stellar mass of
$10^{8.8\pm0.2} ~M_{\odot}$ ($6.3^{+3.7}_{-2.3}\times 10^8 ~M_{\odot}$), while
our optimal bulge mass is $0.84\times10^8 ~M_{\odot}$, or 7.5 times less
massive.

BRGG15 obtained a bulge-to-disk mass ratio $B/D=0.32$ ($B/Total=0.24$) which is
what one typically finds in early-type disk galaxies rather 
than in late-type spiral galaxies (e.g., Laurikainen et al.\ 2010). 
Our bulge-to-disk mass ratio is $\sim$4\%, in good agreement with values 
seen in Scd--Sm galaxies (Graham \& Worley 2008).

\section{Additional parameters and data} 

\subsection{The black hole mass}\label{Sec_BHmass}

Using a `reduced' virial factor $\epsilon = 1$ (which equates to an $f$-factor
equal to 4), BRGG15 reported black hole mass estimates ranging from
(2.7--6.2)$\times10^4 \, f_4 ~M_{\odot}$ when using three different techniques for
modeling the narrow emission lines, plus an associated factor of 2.7
uncertainty arising from the empirical relations used to estimate virial 
black hole masses.  This gives a black hole mass range from $10^4 f_4 ~M_{\odot}$ to 
$1.7 \times 10^5 \, f_4 ~M_{\odot}$ about their adopted nominal value of $5\times 10^4
f_4 ~M_{\odot}$, where $f_4 = f/4$. 

Building on Onken et al.\ (2004) who reported an $f$-factor of 5.5$\pm$1.7,
Graham et al.\ (2011) also investigated the optimal $f$-factor by calibrating
the virial black hole masses obtained from reverberation mapping studies of
AGN against the $M_{\rm bh}$--$\sigma_*$ relation defined by other galaxies with
directly measured black hole masses.  Using both a `forward' and an `inverse' 
linear regression to construct the $M_{\rm bh}$--$\sigma_*$ relation --- the
latter is required for minimizing the sample selection bias that arises from 
the need to spatially resolve the black holes's sphere-of-influence --- 
 --- an $f$-factor of around 3 was derived for the full sample.
Moreover, they reported why this was an upper limit because of several
factors, including radiation pressure (Marconi et al.\ 2008, who also report
$f\approx 3$). Repeating this analysis, the 
forward and inverse linear regressions from Park et al.\ (2012) yielded
similar results to Graham et al.\ (2011), 
and an $f$-factor of 3 is commonly used in the literature (e.g., Kaspi et
al.\ 2000; Xiao et al.\ 2011; Jiang et al.\ 2011).  However when dealing only with AGN
in barred galaxies, Graham et al.\ (2011) found that the optimal 
$f$-factor\footnote{Graham et al.\ (2011) reported optimal $f$-factors of
  $5.4^{+1.5}_{-1.2}$ for unbarred galaxies, $2.3^{+0.9}_{-0.6}$ for barred
  galaxies (many of which may have pseudobulges), and $2.8^{+0.7}_{-0.5}$ for
  the ensemble.} was $2.3^{+0.9}_{-0.6}$ (and this may still be too high, see
Shankar et al.\ 2016).  The optimal virial black hole mass
for LEDA~87300 is thus a factor of 2.3/4 times smaller than reported by
BRGG15, 
and we therefore adopt a mass of $2.9 \times 10^4 f_{2.3} ~M_{\odot}$
with a range from (0.6--9.6)$\times 10^4 f_{2.3} ~M_{\odot}$ ($\log[M_{\rm
    bh}\,f_{2.3}] = 4.46^{+0.52}_{-0.70}$).
This is still consistent with the value adopted by BRGG15. 

Our $M_{\rm bh}/M_{\rm sph,*}$ mass ratio for LEDA~87300 is 0.031\%, which is
2.4 times larger than the value of 0.013\% obtained by BRGG15.  Our ratio is
16 times smaller than the median value of 0.49\% reported by Graham \& Scott
(2013) for massive spheroids with partially depleted cores, and 22 times
smaller than the median value of 0.68\% reported by Savorgnan et al.\ (2015)
for their early-type galaxy sample\footnote{The intrinsic scatter observed in
  these ratios is $\sim$0.4 dex.}.

\subsection{The stellar velocity dispersion}\label{Sec_sigma}

There is no available stellar velocity dispersion ($\sigma_*$) for
LEDA~87300.  However BRGG15 measured a gas velocity dispersion of
$27\pm5$ km s$^{-1}$ from the [NII] line.  They assumed 0.15 dex of
scatter and zero mean offset between measurements of the stellar and
the gas velocity dispersion in other galaxies, and reported $\sigma_*
= 27^{+12}_{-10}$ km s$^{-1}$.  There is however a well known tendency
for gas to have a lower velocity dispersion than stars (e.g., Vega
Beltr\'an et al.\ 2001, their Figure~3; Barth et al.\ 2008, their
Figure~10), with Ho (2009) reporting an average $\sigma_{\rm
  [NII]}/\sigma_*$ ratio of 0.8.  That is, the stellar velocity
dispersion is typically 25\% higher than the gas velocity
dispersion. Although, this ratio is less clear for the Scd-Sm galaxies
for which there is more scatter, with Figure~1 from Ho (2009)
revealing that systems with $\sigma_{\rm [NII]} \approx 25$--35 km
s$^{-1}$ have $\sigma_* \approx 23$--100 km s$^{-1}$.  Figure~11d from
Xiao et al.\ (2011) reveals this same behavior at low velocity
dispersions, and that one expects $\sigma_* \approx$ 30--50 km
s$^{-1}$ if $\sigma_{\rm [NII]} = 27$ km s$^{-1}$.  We adopt this more
constrained range rather than an average value of 50--60 km s$^{-1}$
which might be expected from Ho (2009).  
Adding the $\pm5$ km s$^{-1}$ measurement uncertainty in
quadrature with a $\pm10$ km s$^{-1}$ range, we conservatively adopt
$\sigma_* = 40 \pm 11$ km s$^{-1}$.

Given that spheroids with low S\'ersic indices have rather flat,
luminosity-weighted, velocity dispersion profiles (e.g., Graham \& Colless
1997, their Figure~7), no discernible difference between the central
($R\approx0$) and effective ($R \le R_{\rm e}$) aperture velocity dispersion
is expected for the spheroidal component of LEDA~87300.  Of course the
presence of discs and bars can bias attempts to measure the spheroid's
velocity dispersion, and increasingly so as one moves to apertures of larger
radii in disk galaxies.  This is because one will increasingly acquire more of
a galaxy velocity dispersion than a bulge velocity dispersion as the radial
range is increased.  For this reason, one should be wary of $\sigma_{R_{\rm
    e}}$ or $\sigma_{R_{\rm e}/2}$, which may be systematically biased if the
disc light and dynamics contribute more in galaxies with lower mass black
holes.

\subsection{The X-ray Data}\label{Sec_X}

\begin{figure}
\begin{center}
\includegraphics[trim=2cm 0cm 2cm 0cm, width=0.49\columnwidth]{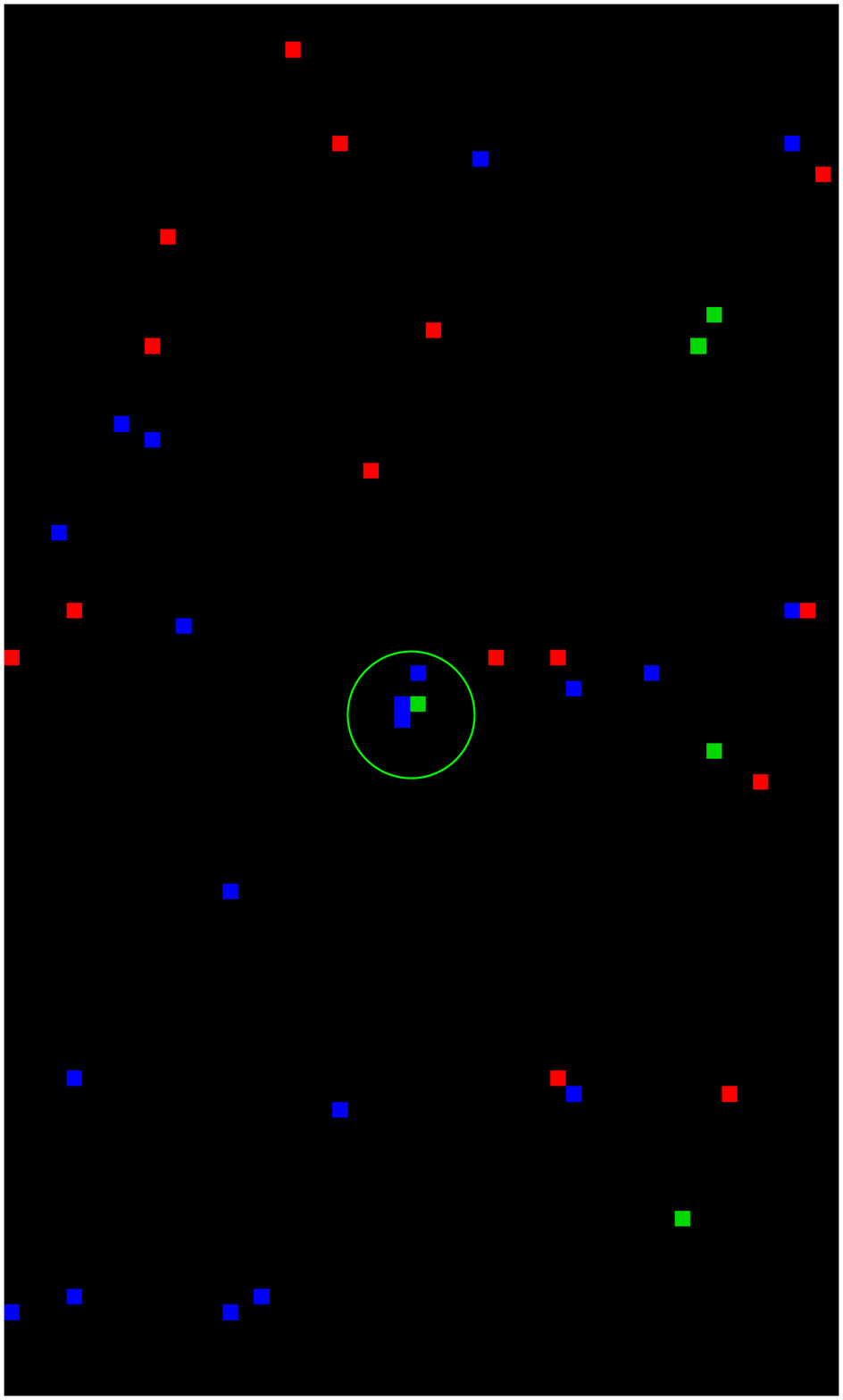}
\includegraphics[trim=2cm 0cm 2cm 0cm, width=0.49\columnwidth]{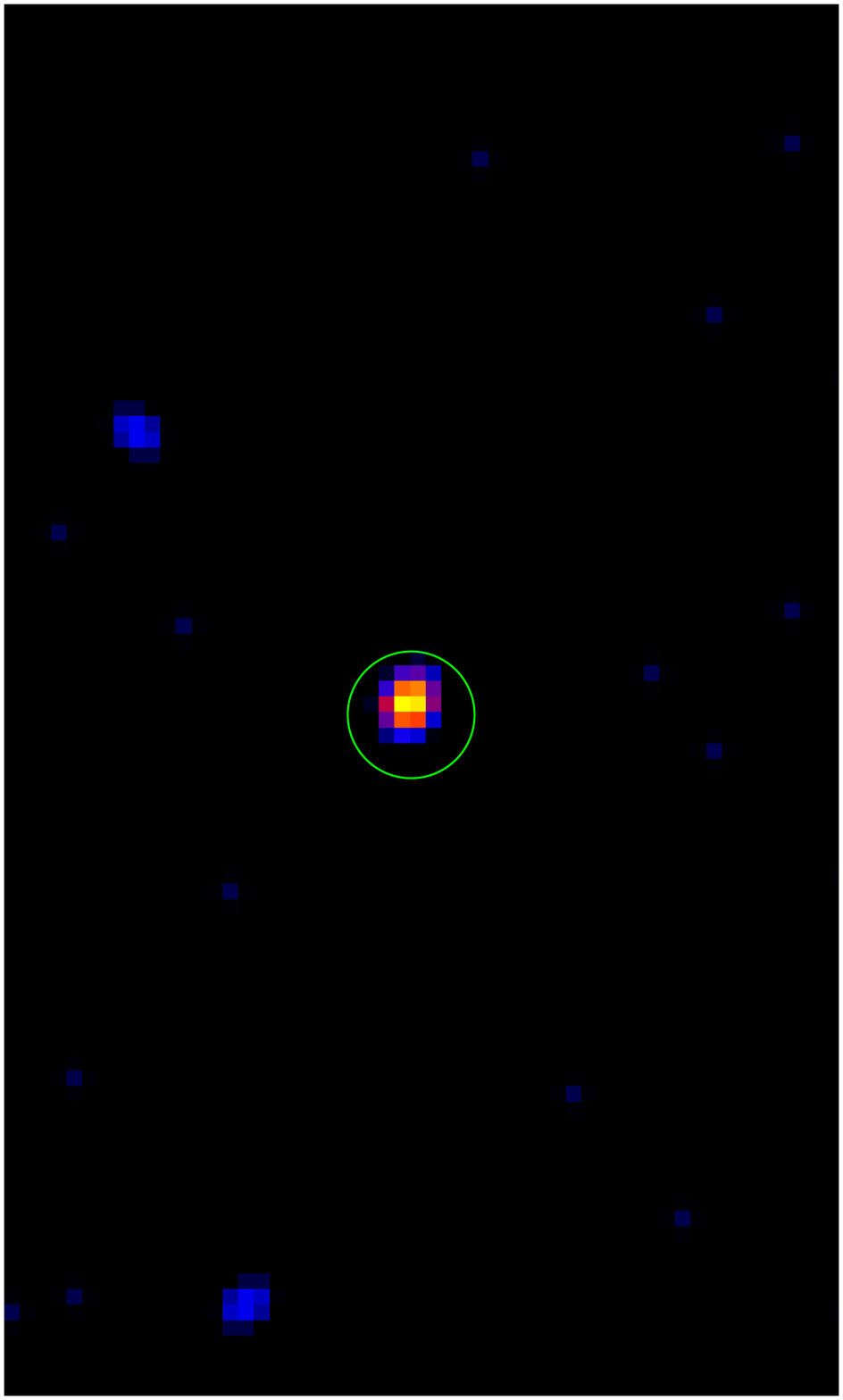}
\caption{
Left panel: {\it Chandra X-ray Observatory} ACIS-S photon image such that 
red=0.3--1.0 keV, green=1--2 keV and blue=2--7 keV. 
The green point within the green circle of radius equal to 2$\arcsec$ centered
on LEDA~87300 has an energy between 1.5 and 2 keV,
while the 3 blue points within the circle have an energy between 2 and 5 keV. 
Right panel: The same image after smoothing with a Gaussian having $\sigma =
3$ pixels.  East is up and North is to the right. 
}
\label{Fig7}
\end{center}
\end{figure}

Given the current scarcity, and thus importance, of IMBHs, we have
additionally re-analyzed the {\it Chandra X-ray Observatory} Advanced
CCD Imaging Spectrometer (ACIS) data for LEDA~87300.  We provide
confirmation of the analysis in BRGG15, and therefore we do not
discuss this data at any great length.  Within the 5.5 hour exposure,
4 X-ray photons were detected within a circle of radius 2$\arcsec$
centered on the galaxy (see Figure~\ref{Fig7}).  Given the average
background count was just 0.25 for an area of this size, we have a
90\% confidence limit of between 1 and 8 photons (Kraft et al.\ 1991).
That is, we have {\it at least} a 90\% detection of a point-like X-ray
source in LEDA~87300.  We acknowledge that ``point-like'' actually
means within $\sim$1$\arcsec$ or $\sim$500 pc, and thus the source of
the X-rays could be an AGN, a compact starburst nucleus, or a
combination of both. 

None of the four photons appeared in the 0.3--1.5 keV band, they all had
energies in the 1.5--5 keV range, with just one having an energy between 1.5
and 2.0 keV.  It is interesting that there are no photons with energies in the
0.3--1.5 keV band because it requires the soft band to be absorbed by a column
density of neutral hydrogen of at least $5\times10^{21}$ and more likely
$10^{22}$ cm$^{-2}$.  Assuming a plausible photon index\footnote{The photon
  index is defined as $dN/dE \propto E^{-\Gamma}$ where $N$ is the number flux
  of photons, e.g.\ Ishibashi \& Courvoisier (2010).} for the slope of the
X-ray spectrum somewhere in the range $\Gamma \approx 1.4$--$1.7$, one
detected photon corresponds to a 2--10 keV luminosity $L_X \approx 10^{39}$
erg s$^{-1}$. As noted by BRGG15, the bolometric luminosity is therefore
$\approx 4\times 10^{40}$ erg s$^{-1}$ if using $L_{\rm bol}/L_{2-10}
 \sim 10$ (Marconi et al.\ 2004; Vasudevan \& Fabian 2007). 

The rest of the galaxy does not appear to contribute to the X-ray emission,
and is not expected to.  For a total galaxy stellar mass of $\sim$$4\times10^9
~M_{\odot}$, the expected X-ray luminosity from the low-mass X-ray binaries
(LMXBs) is $L_X \approx (8\pm1)\times10^{28}\,M_{\rm galaxy,*} ~ \approx ~
3\times10^{38}$ erg s$^{-1}$ (Gilfanov 2004).  In the {\it Chandra} exposure,
this would correspond to only 1/3 of a photon in 5.5 hours.  An additional
component may come from high-mass X-ray binaries (HMXBs), however if the star
formation rate is less than $\sim$0.5 $M_{\odot}$ yr$^{-1}$, then we do not
expect HMXBs to contribute significantly.

\begin{figure*}
\begin{center}
\includegraphics[trim=4cm 2.3cm 3cm 9.5cm, width=\textwidth]{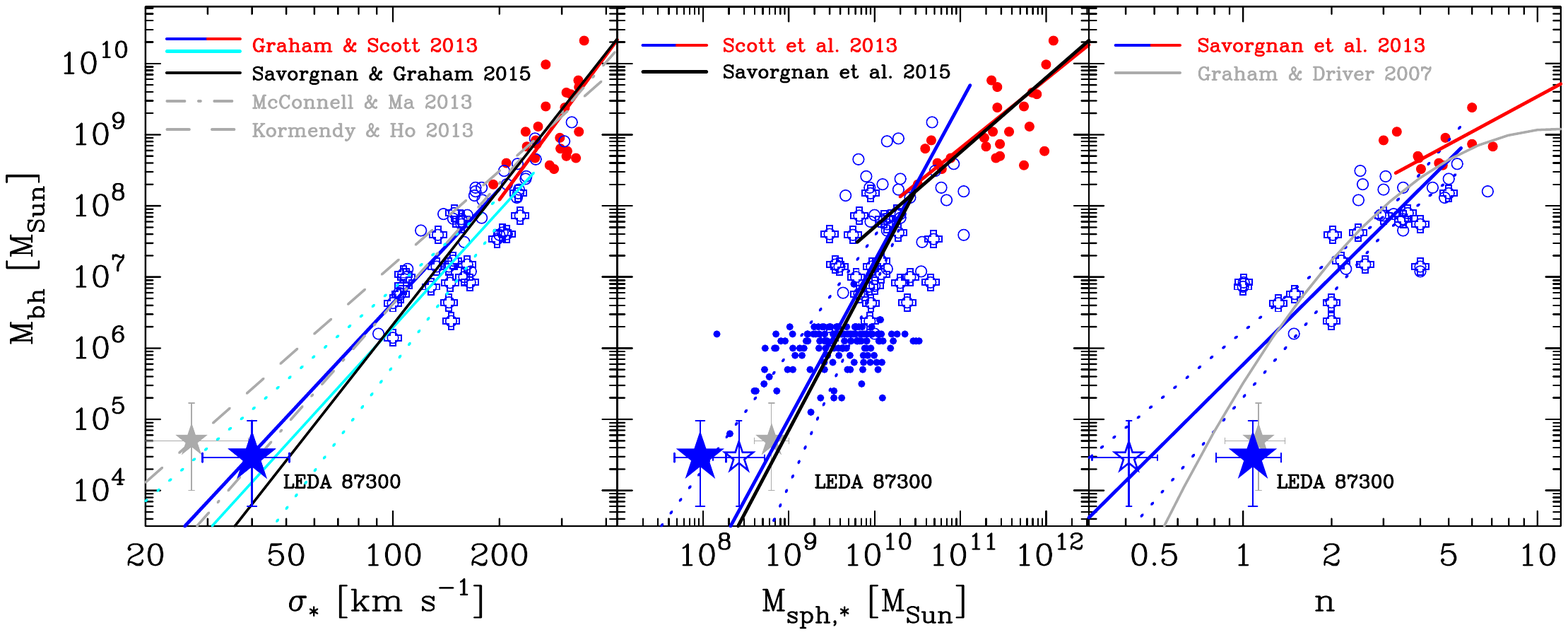}
\caption{
Left panel: $M_{\rm bh}$--$\sigma_*$ data and relations from Graham \& Scott
(2013): core-S\'ersic galaxies (red dots and red line); 
unbarred S\'ersic galaxies (blue circles and blue line); 
barred galaxies (blue crosses and cyan line). 
The dotted cyan lines delineate the 1-sigma uncertainty on the latter relation. 
The gray star shows the parameters for LEDA~87300 from 
BRGG15, while the  blue star shows our parameters.  
Note: Kormendy \& Ho (2013) adjusted the velocity dispersions to create 
reduced velocity dispersion estimates within 
$R_{\rm e}/2$ ($\sigma_{*,R_{\rm e}/2}$) to which 
their $M_{\rm bh}$--$\sigma_*$ line pertains. 
Middle panel: 
$M_{\rm bh}$--$M_{\rm sph,*}$ diagram and relations from Scott
et al.\ (2013): core-S\'ersic galaxies (red); 
barred and unbarred S\'ersic galaxies (blue).
The AGN shown here as small blue dots come from Graham \& Scott (2015) but
were not used to derive the blue line. 
The dotted blue lines delineate the 1-sigma uncertainty of the solid blue line.  
The gray star shows the ``barge'' (bar + bulge) parameters for LEDA~87300 from
BRGG15, while the open blue star shows our parameters for the barge. 
The larger solid blue star shows our parameters for the bulge of LEDA~87300. 
Right panel: $M_{\rm bh}$--$n$ data and relations from Savorgnan et 
al.\ (2013).
} 
\label{Fig8}
\end{center}
\end{figure*}

\section{Discussion}\label{Sec_DC}

BRGG15 reported an ellipticity of 0.63 for their ``bulge'' 
component of LEDA~87300 and wrote that this value is in accord with expectations
for pseudobulges.  However because their ``bulge'' model was dominated by
the bar (and perhaps the start of the spiral arms), they
over-estimated the ellipticity of the actual spheroid which likely has
an ellipticity half of this value (see Figure~\ref{Fig6}).  As
discussed in Graham (2014), pseudobulges are particularly
difficult to identify, and as such we do not label the bulge component
of LEDA~87300 one way or another.

Figure~\ref{Fig8} shows the location of LEDA~87300 in the $M_{\rm
  bh}$--$\sigma_*$, $M_{\rm bh}$--$M_{\rm sph,*}$, and $M_{\rm bh}$--$n$
diagrams.  We have used the virial black hole mass estimated with the optimal
virial factor for barred galaxies (section~\ref{Sec_BHmass}), and we now
discuss each panel in turn, starting with the $M_{\rm bh}$--$\sigma_*$
diagram.

BRGG15 showed that their black hole mass and adopted stellar velocity
dispersion for LEDA~87300 placed it on the extrapolation of the $M_{\rm
  bh}$--$\sigma_*$ relation from Kormendy \& Ho (2013) which was constructed
using a sample which Kormendy \& Ho (2013) considered not to be
`pseudobulges'.  In the left panel of Figure~\ref{Fig8}, one can see that our
slightly revised location of LEDA~87300 places it on the $M_{\rm
  bh}$--$\sigma_*$ relation from McConnell \& Ma (2013) for barred and
unbarred galaxies combined.  LEDA~87300 additionally agrees well with the $M_{\rm
  bh}$--$\sigma_*$ relations from Graham \& Scott (2013) for both barred and
unbarred galaxies, which are themselves consistent with the relations
presented in Graham et al.\ (2011). 

Graham (2008) and Hu (2008) showed that barred galaxies --- 
which may contain pseudobulges, classical bulges, or simultaneously both a classical
bulge and a pseudobulge (e.g., Erwin et al.\ 2003) --- have a tendency to
sometimes be offset from the $M_{\rm bh}$--$\sigma_*$ relation in the
direction of higher velocity dispersions, as explained by Hartmann et
al.\ (2014).  It was pointed out in section~\ref{Sec_sigma} that the stellar velocity
dispersion may be $\approx$50-60 km s$^{-1}$. If correct, 
LEDA~87300 would remain within the 1-sigma 
uncertainty of the $M_{\rm bh}$--$\sigma_*$ relation for barred galaxies
shown in Figure~\ref{Fig8}. 
The uncertainties associated with the black hole mass and stellar velocity dispersion 
for LEDA~87300, coupled with 
the uncertainty on the slope and intercept of the 
various $M_{\rm bh}$--$\sigma_*$ scaling 
relations, are such that we are not advocating any one relation
over another, but expanding upon the consistency noted by BRGG15 between
LEDA~87300 and the extrapolation of the $M_{\rm bh}$--$\sigma_*$ relations to
lower velocity dispersions. 
Finally, we note that these relations may yet turn out to be upper envelopes
(Batcheldor 2010), with Subramanian et al.\ (2015) presenting evidence for this
with a sample of low surface brightness galaxies for which the black hole
masses, estimated using the broad H$\alpha$ line, fell well below the various
$M_{\rm bh}$--$\sigma_*$ relations. 

Including over 100 AGN with virial black hole mass estimates from Jiang et
al.\ (2011), Graham \& Scott (2015) extended the $M_{\rm bh}$--$M_{\rm sph,*}$
diagram from $M_{\rm bh} = 10^6 ~M_{\odot}$ down to $10^5 ~M_{\odot}$. They
revealed that the bulges of these AGN follow the near-quadratic relation
presented in Scott et al.\ (2013) for galaxies without partially depleted
cores, i.e.\ for classical spheroids and pseudobulges alike.  Mezcua et
al.\ (2015b) present additional AGN supporting such a steep relation,
bolstering support for theoretical models which predict a steep scaling
relation at these masses (e.g., Fontanot et al.\ 2006, 2015; 
Dubois et al.\ 2012; Bonoli et al.\ 2014; Bellovary et al.\ 2014). Building
on BRGG15, we extend the $M_{\rm bh}$--$M_{\rm sph,*}$ diagram down to $M_{\rm
  bh} \sim 3\times10^4 ~M_{\odot}$ through the inclusion of LEDA~87300
(Figure~\ref{Fig8}, middle panel).  The difference here is that we have used
the (dust-corrected) bulge mass rather than a combination of the bulge plus
bar mass.  This had the effect of shifting LEDA~87300 to the left in this
diagram.  Although the previously reported location of LEDA~87300
(Figure~\ref{Fig1}) better matches the near-quadratic $M_{\rm bh} \propto
M_{\rm sph,*}^{2.22\pm0.58}$ relation from Scott et al.\ (2013), within the
1-sigma uncertainties there remains agreement between LEDA~87300 and this
steep relation.  Lastly, we note that while the use of the barge or galaxy
luminosity will spread the data to the right in the $M_{\rm bh}$--$M_{\rm
  sph,*}$ diagram, star formation activity may also produce over-luminous
bulges, as discussed by Busch et al.\ (2015).

LEDA~87300 can be seen to roughly follow the observed trend in the $M_{\rm
  bh}$--$n$ diagram (Figure~\ref{Fig8}, right panel).  
However LEDA~87300 is offset below the linear $M_{\rm bh}$--$n$ relation. 
For a S\'ersic index 
$n_{\rm maj}=1.08$, and a typical 25\% uncertainty\footnote{Equation~3
  from Graham \& Driver (2007) reveals how this uncertainty propagates through
  to the predicted black hole mass. An intrinsic scatter of 0.3 dex in the
  $\log(M_{\rm bh})$ direction has been used here.}, the log-linear relation
from Savorgnan et al.\ (2013) is such that
$\log(M_{\rm bh}/M_{\odot}) = (7.73\pm0.12) + (4.11\pm0.72)\log(n/3)$, and 
gives $\log(M_{\rm bh}/M_{\odot})=5.91\pm 0.64$, or equivalently $M_{\rm
  bh} = 8.1^{+27.4}_{-6.2} \times 10^5 ~M_{\odot}$.  The lower 1-sigma
bound to this mass range ($1.9\times 10^5 ~M_{\odot}$) is two times higher 
than our upper range 
to the virial black hole mass (0.96$\times 10^5 \, f_{2.3} ~M_{\odot}$).  
The curved $M_{\rm bh}$--$n$ relation from Graham \& Driver (2007) does 
better, but this is not our preferred solution because it does not separate
galaxies into core-S\'ersic versus S\'ersic galaxies, or early-type versus
late-type galaxies.  

It should be born in mind that LEDA~87300 is just one data point, and the
scaling relations are still being refined.  Close agreements today may not be
close agreements tomorrow, and thus one should be mindful of the uncertainty
associated with the relations and the data.  More IMBH data would be welcome, 
In this regard, while Reines et al.\ (2013) have 
identified candidate IMBHs in dwarf galaxies, and Sartori et al.\ (2015) are
pursuing AGN in low-mass galaxies, 
Graham \& Scott (2013) have 
identified targets in the small bulges of disk galaxies.  Graham \& Scott
(2013) named 41 candidate galaxies for hosting an intermediate mass black hole
(IMBH) based on the spheroid $K$-band magnitudes reported in the Dong \& De
Robertis (2006) study of galaxies with low-luminosity AGN.  These spheroid
magnitudes are being re-derived (Ciambur et al., in preparation) with new
images having better spatial resolution than the $2\arcsec .5$ Two Micron All
Sky Survey (2MASS) data\footnote{http://www.ipac.caltech.edu/2mass} images
used by Dong \& De Robertis (2006).  This will provide improved spheroid
magnitudes and thus better black hole mass estimates.  An observational
campaign (Webb et al., in preparation) to collect 5 GHz radio data and X-ray
data will provide an independent means for estimating/confirming these black
hole masses via the `fundamental plane of black hole activity' (Merloni et
al.\ 2003; Falcke et al.\ 2004; Dong \& Wu 2015; Liu et al.\ 2015; Nisbet \&
Best 2016) and
hopefully enable the further population of the black hole scaling diagrams at
masses below $10^5 ~M_{\odot}$.

\acknowledgments

This research was supported under the Australian Research Council’s
funding scheme (FT110100263). 
This research has made use of the NASA/IPAC Extragalactic Database (NED). 
Funding for {\it SDSS}-III has been provided by the Alfred P.\ Sloan Foundation, the
Participating Institutions, the National Science Foundation, and the
U.S.\ Department of Energy Office of Science.

\end{document}